\documentclass[12pt, a4paper]{article}
\usepackage{amsmath,amssymb,bm,epsfig,afterpage}
\usepackage{cite}
\usepackage{here}
\usepackage{color}
\usepackage{colortbl}
\usepackage{xcolor}
\usepackage{tabularx}
\usepackage{subcaption}
\usepackage{bbm}
\usepackage{graphicx}
\usepackage{listings}
\usepackage{longtable}
\usepackage[utf8]{inputenc}

\makeatletter
    
    \@addtoreset{equation}{section}
  \makeatother

\definecolor{Orange}{cmyk}{0,0.61,0.87,0}
\definecolor{JungleGreen}{cmyk}{0.99,0,0.52,0}
\definecolor{OliveGreen}{cmyk}{0.64,0,0.95,0.40}
\definecolor{Brown}{cmyk}{0,0.81,1,0.60}
\definecolor{RoyalBlue}{cmyk}{0.71,0.53,0,0.12}
\definecolor{Gray}{cmyk}{0,0,0,0.40}
\definecolor{LightPink}{cmyk}{0.0,0.25,0,0}
\definecolor{LLightPink}{cmyk}{0.0,0.10,0,0}
\definecolor{LightBlue}{cmyk}{0.25,0,0,0}
\definecolor{LightGray}{cmyk}{0,0,0,0.2}

\newcommand{\vev}[1]{{\langle{#1}\rangle}}
\newcommand{\br}[2]{\text{Br}({#1}\to{#2})}

\newcommand{{\Lcal}}{\mathcal{L}}
\newcommand{\Mcal}{\mathcal{M}}

\newcommand{\Hcal}{\mathcal{H}}
\newcommand{\Ocal}{\mathcal{O}}

\newcommand{\ket}[1]{|{#1}\rangle}
\newcommand{\bra}[1]{\langle{#1}|}

\newcommand{\re}[1]{\text{Re} \left( {#1} \right)}
\newcommand{\abs}[1]{\left|{#1}\right|}
\newcommand{\Br}[2]{\text{Br}\left({#1}\to{#2}\right)}

\newcommand{\ol}[1]{\overline{#1}}
\newcommand{\ti}[1]{\tilde{#1}}

\newcommand{\hg}{\hat{g}}
\newcommand{\hY}{\hat{Y}}

\newcommand{\he}{\hat{e}}

\newcommand{\Pfb}{P_{\ol{5}}}

\newcommand{\eps}{\varepsilon}
\newcommand{\la}{\lambda}

\newcommand{\gam}{\gamma}

\newcommand{\U}[1]{\ensuremath{\mathrm{U}(#1)}}

\newcommand{\SU}[1]{\ensuremath{\mathrm{SU}(#1)}}

\newcommand{\bsll}{b\to s\ell^+\ell^-}
\newcommand{\zv}{\boldsymbol{z}}
\newcommand{\nv}{\boldsymbol{n}}
\newcommand{\zerov}{\boldsymbol{0}}

\allowdisplaybreaks[3]
\newcolumntype{Y}{&gt;{\centering\arraybackslash}X}
\usepackage[colorlinks=true, linkcolor=OliveGreen, citecolor=RoyalBlue,
urlcolor=RoyalBlue]{hyperref}

\definecolor{darkgreen}{HTML}{109930}

\begin{document}

\begin{titlepage}

\begin{flushright}
{\tt
}
\end{flushright}

\vskip 1.35cm
\begin{center}

{\Large
{\bf
Complete Vector-like Fourth Family and new \texorpdfstring{$\boldsymbol{\U1'}$}{\U1} for Muon Anomalies
}
}

\vskip 1.5cm

Junichiro~Kawamura$^{a,b,}$\footnote{%
\href{mailto:kawamura.14@osu.edu}{\tt kawamura.14@osu.edu}},
Stuart~Raby$^{a,}$\footnote{%
\href{mailto:raby.1@osu.edu}{\tt raby.1@osu.edu}},
and
Andreas Trautner$^{c,}$\footnote{%
\href{mailto: trautner@mpi-hd.mpg.de}{\tt trautner@mpi-hd.mpg.de}}
\vskip 0.8cm

{\it $^a$Department of Physics, Ohio State University, Columbus, Ohio
 43210, USA}
\\[3pt]
{\it $^b$Department of Physics, Keio University, Yokohama 223-8522, Japan}
\\[3pt]
{\it $^c$Max-Planck-Institut f$\ddot{\text{u}}$r Kernphysik, Saupfercheckweg 1, 69117 Heidelberg, Germany}
\\[3pt]

\date{\today}

\vskip 1.5cm

\begin{abstract} 
We consider the Standard Model (SM) with the addition of a $\mathrm{U(1)^\prime}$ 
gauge symmetry and a complete fourth family of quarks and leptons which are vector-like with respect to 
the full $\mathrm{SU(3)_C}\times \mathrm{SU(2)_L} \times \mathrm{U(1)_Y}\times \mathrm{U(1)^\prime}$ gauge symmetry.   
The model provides a unified explanation of experimental anomalies in $(g - 2)_\mu$ as well as $b \rightarrow s \ell^+ \ell^-$ decays.
We find good fits to the
deviations from the SM, while at the same time fitting all other SM observables. 
The model includes a new $Z^\prime$ gauge boson, a $\mathrm{U(1)^\prime}$-breaking scalar, and vector-like leptons all with mass of order a few $100$ GeV.  
It is consistent with all currently released high energy experimental data, 
however, it appears imminently testable with well designed future searches.
Also precision flavor experiments, especially more accurate direct determinations of CKM matrix elements,
would allow to probe the best fit points.
\end{abstract}

\end{center}

\end{titlepage}
\setcounter{footnote}{0}

\section{Introduction}
The Standard Model (SM) is very successful,
explaining most experimental results.
However, there are experimental discrepancies with some SM predictions.
One is the anomalous magnetic moment of the muon, $(g-2)_\mu$.
The current experimental measurement~\cite{Tanabashi:2018oca,Bennett:2006fi}
shows the discrepancy,
\begin{align}
 \Delta a_\mu := a_\mu^\text{exp} - a_\mu^\text{SM} = 268 (63) (43) \times 10^{-11}.
\end{align}
Other discrepancies are reported in observables related to
$\bsll$ processes.
The observables testing lepton flavor non-universality, $R_K$ and $R_{K*}$,
deviate from the SM prediction~\cite{Aaij:2014ora,Aaij:2017vbb},
even though the most recent data is consistent with the SM at $2.5\,\sigma$~\cite{Aaij:2019wad}
or has large error bars \cite{Abdesselam:2019wac}. There are also deviations from the SM predictions
for semi-leptonic branching ratios~\cite{Aaij:2013aln,Lees:2013nxa,Aaij:2014pli,Aaij:2015esa}
and angular distributions~\cite{Aaij:2013qta,Khachatryan:2015isa,Aaij:2015oid,Abdesselam:2016llu,Wehle:2016yoi,ATLAS:2017dlm,CMS:2017ivg}.

It is interesting that all the observables for $\bsll$ can be explained
by New Physics (NP) contributions to
the effective Hamiltonian~\cite{Buras:1994dj,Bobeth:1999mk},
\begin{align}
\label{eq-defC9C10}
\Hcal_\text{eff}^{\ell} = -\frac{4G_F}{\sqrt{2}}
                    \frac{\alpha_e}{4\pi} V_{tb}V^*_{ts} \sum_{a=9,10}
                     \left(C_a^{\ell} \Ocal_a^{\ell}+C_a^{\prime\ell} \Ocal_a^{\prime\ell}  \right),
\end{align}
where $\ell = e, \mu, \tau$ and
the operators are defined as
\begin{align}
 \Ocal_9^{\ell} :=&\   \left[ \ol{s} \gam^\mu P_L b  \right]
                                                             \left[ \ol{\ell} \gam_\mu \ell \right],\quad
 \Ocal_{10}^{\ell} :=  \left[ \ol{s} \gam^\mu P_L b  \right]
                                                             \left[ \ol{\ell} \gam_\mu \gamma_5 \ell \right], \\
 \Ocal_9^{\prime\ell} :=&\   \left[ \ol{s} \gam^\mu P_R b  \right]
                                                             \left[ \ol{\ell} \gam_\mu \ell \right],\quad
 \Ocal_{10}^{\prime\ell} :=   \left[ \ol{s} \gam^\mu P_R b  \right]
                                                             \left[ \ol{\ell} \gam_\mu \gamma_5 \ell \right].
\end{align}
The analyses before Moriond 2019~\cite{
Altmannshofer:2017fio,Altmannshofer:2017yso,Alok:2017sui,Capdevila:2017bsm,Ciuchini:2017mik,DAmico:2017mtc,Geng:2017svp,Ghosh:2017ber,Arbey:2018ics}
and after Moriond 2019~\cite{Aebischer:2019mlg,Alguero:2019ptt,Alok:2019ufo,Ciuchini:2019usw,Datta:2019zca,Kowalska:2019ley,Arbey:2019duh,Kumar:2019nfv}
show that several patterns of NP contributions explain the discrepancies significantly better than the SM.
In all cases, there should be a sizable negative contribution to $C_9^\mu$.

The muon anomalous magnetic moment $\Delta a_\mu$
can be explained by introducing vector-like (VL) leptons which exclusively couple
to muons~\cite{Czarnecki:2001pv,Kannike:2011ng,Dermisek:2013gta,Lindner:2016bgg}.
A way to address the $\bsll$ anomalies is to introduce a $Z'$ gauge boson
which couples to muons and down-type quarks.
For instance, $\U1_{\mu\text{-}\tau}$ gauge symmetry and VL quarks (with $\U1_{\mu\text{-}\tau}$ charge) are introduced
to control the flavor dependent couplings of the $Z'$ boson~\cite{Altmannshofer:2014cfa,Crivellin:2015mga}
(see Ref.~\cite{King:2017anf} for a more general discussion of $\U1'$ gauge symmetry).
It is shown in Refs.~\cite{Allanach:2015gkd,Altmannshofer:2016oaq,Megias:2017dzd,Raby:2017igl,Darme:2018hqg}
that both of $\Delta a_\mu$ and $\bsll$ anomalies are successfully explained
in models with VL-leptons, VL-quarks and a $Z'$ boson.

Box-diagram contributions involving new fermions and scalars can also account
for the $\bsll$ anomalies~\cite{Gripaios:2015gra,Arnan:2016cpy,Grinstein:2018fgb,Arnan:2019uhr}. They may even include
particle candidates for dark matter~\cite{Chiang:2017zkh,Cline:2017qqu,Kawamura:2017ecz,Barman:2018jhz,Cerdeno:2019vpd}.
These extensions, however, typically induce deviations from the SM predictions
for Lepton Flavor Violating (LFV) decays, Higgs decays, and $B_s\text{-}\ol{B}_s$ mixing.

In this paper, we propose a model with a complete fourth family of fermions
which are VL under both the SM and a $\U1'$ gauge symmetry.
Similar models with chiral $\U1'$ gauge symmetry were considered
in Refs.~\cite{Altmannshofer:2014cfa, Raby:2017igl}.
In these models, the SM families typically have sizable couplings to the $Z'$ gauge boson in the gauge basis.
A VL $\U1'$ has been studied where a new singlet scalar~\cite{Sierra:2015fma} 
or the singlet VL neutrino~\cite{Falkowski:2018dsl} are dark matter candidates. 
However, the parameter space there is very restricted, so that $\Delta a_\mu$ was not addressed.
In our present model, all $Z'$ couplings to the SM fermions are controlled
by mixing of the SM families with the VL family.
We find that a certain pattern of mixings can simultaneously
address both, $\Delta a_\mu$ and the $\bsll$ anomalies.

We analyze this model involving all three SM families.
This allows us to explicitly discuss both the CKM matrix
and exotic particle production from quarks and gluons at the Large Hadron Collider (LHC).
We find points in the parameter space which explain the muon anomalies
and all other observables by using a $\chi^2$ fit.
The purpose of this paper is to demonstrate the existence of points
which are consistent with the anomalies, as well as all other SM observables,
and study the expected phenomenology at these points.
A more detailed analysis of the expected phenomenology in a wider parameter space
is delegated to future work.

The rest of this paper is organized as follows.
The model is introduced in Section~\ref{sec-model},
then we discuss the most relevant observables for the muon anomalies in Section~\ref{sec-obs}.
In Section~\ref{sec-rslt} we show the best fit points of our $\chi^2$ analysis and study their phenomenology.
Section~\ref{sec-concl} is devoted to our conclusions.
Values of the input parameters and all observables calculated
in this analysis are listed in the Appendix.

\section{Model}
\label{sec-model}

\subsection{Matter Content and Masses}
\begin{table}
\centering
\caption{\label{tab-SM}
Quantum numbers of SM particles. Here, $i=1,2,3$ runs over the three SM families.
Electromagnetic charges are given by $Q_f = T_f^3 + Y_f/2$. 
}
 \begin{tabular}{c|cccccc|c}\hline
                  & ${q_L}_i$&${\ol{u}_R}_i$&${\ol{d}_R}_i$&${l_L}_i$&${\ol{e}_R}_i$
                  &${\ol{\nu}_R}_i$&$H$ \\ \hline\hline
 $\SU3_{\mathrm{C}}$&$\bf{3}$&$\ol{\bf{3}}$& $\ol{\bf{3}}$&$\bf{1}$&$\bf{1}$& $\bf{1}$&$\bf{1}$\\
 $\SU2_{\mathrm{L}}$&$\bf{2}$&$\bf{1}$& $\bf{1}$& $\bf{2}$& $\bf{1}$& $\bf{1}$&$\bf{2}$ \\
 $\U1_{\mathrm{Y}}$ &$1/3$&$\text{-}4/3$&$2/3$&$\text{-}1$&$2$&$0$&$\text{-}1$ \\ \hline
 $\U1'$     & 0&0&0&0&0&0&0 \\ \hline
 \end{tabular} \\
\vspace{0.5cm}
\caption{\label{tab-extra}
Quantum numbers of the new VL fourth family and new scalar fields.
}
 \begin{tabular}{c|cccccc|cccccc|cc}\hline
         & $Q_L$&$\ol{U}_R$&$\ol{D}_R$&$L_L$&$\ol{E}_R$&$\ol{N}_R$&
         $\ol{Q}_R$&$U_L$&$D_L$&$\ol{L}_R$&$E_L$&$N_L$&
         $\phi$ &$\Phi$\\ \hline\hline
 $\SU3_{\mathrm{C}}$&$\bf{3}$&$\ol{\bf{3}}$& $\ol{\bf{3}}$&$\bf{1}$&$\bf{1}$& $\bf{1}$&
                      $\ol{\bf{3}}$&$\bf{3}$& $\bf{3}$&$\bf{1}$&$\bf{1}$& $\bf{1}$&$\bf{1}$&$\bf{1}$\\
 $\SU2_{\mathrm{L}}$&$\bf{2}$&$\bf{1}$& $\bf{1}$& $\bf{2}$& $\bf{1}$& $\bf{1}$&
                      $\bf{2}$&$\bf{1}$& $\bf{1}$& $\bf{2}$& $\bf{1}$& $\bf{1}$&$\bf{1}$&$\bf{1}$   \\
 $\U1_{\mathrm{Y}}$ &$1/3$&$\text{-}4/3$&$2/3$&$\text{-}1$&$2$&$0$&
                    $\text{-}1/3$&$4/3$&$\text{-}2/3$&$1$&$\text{-}2$&$0$&
                    $0$&$0$ \\ \hline
 $\U1'$     & $\text{-}1$ & $1$ & $1$ & $\text{-}1$ & $1$ & $1$ & $1$ & $\text{-}1$ & $\text{-}1$ & $1$ & $\text{-}1$ & $\text{-}1$ & $0$ & $\text{-}1$ \\ \hline
 \end{tabular}
\end{table}

In this paper,
we study a model with a complete VL fourth family and $\U1'$ gauge symmetry.
The quantum numbers of all fields are listed in Tables~\ref{tab-SM} and~\ref{tab-extra}.
The SM $\SU2_{\mathrm{L}}$ doublets are defined as
${q_L}_i = ({u_L}_i, {d_L}_i)$, ${l_L}_i=({\nu_L}_i, {e_L}_i)$, and $H=(H_0, H_-)$.
The new doublets are $Q_L = (U'_L, D'_L)$, $L_L = (N'_L, E'_L)$, $\ol{Q}_R = (-\ol{D}'_R, \ol{U}'_R)$, and $\ol{L}_R = (-\ol{E}'_R, \ol{N}'_R)$.
The model is trivially anomaly-free since the $\U1'$ charge assignment is completely vector-like.

In the gauge basis, there are no couplings between the SM families and the $Z'$ boson.
These are induced in the mass basis by mixing effects.
The Yukawa couplings in the gauge basis are given by
\begin{align}
\Lcal_\mathrm{Yukawa} =&\ \Lcal_\mathrm{SM} +\Lcal_\mathrm{H} +\Lcal_\phi +\Lcal_\Phi+\mathrm{h.c.}\;,
\end{align}
where
\begin{align}
\label{eq-yukSM}
\Lcal_\mathrm{SM} :=&\
          {\ol{u}_R}_i y^u_{ij} {q_L}_{j}\ti{H}  +{\ol{d}_R}_i y^d_{ij} {q_L}_j H +{\ol{e}_R}_i y^e_{ij} {l_L}_j H
                    + {\ol{\nu}_{R}}_i  y^n_{ij} {l_L}_j \ti{H}\;, \\
\Lcal_\mathrm{H}    :=&\ \la_{u} \ol{U}_R Q_L \ti{H} + \la_{d} \ol{D}_R Q_L H+ \la_{e} \ol{E}_R L_L H
                          + \la_{n}  \ol{N}_R L_L \ti{H}  \notag \\
               &\ + \la'_{u} \ol{Q}_R H U_L - \la'_{d} \ol{Q}_R \ti{H} D_L - \la'_{e} \ol{L}_R \ti{H} E_L
                          + \la'_{n}  \ol{L}_R H N_L\;, \\
\Lcal_\phi  :=&\ \phi \left( \la^Q_{V} \ol{Q}_R Q_L-\la^U_{V} \ol{U}_R U_L-\la^D_{V} \ol{D}_R D_L \right. \notag\\
               &\ + \left.\la^L_{V} \ol{L}_R L_L-\la^E_{V}\ol{E}_R E_L-\la^N_{V}\ol{N}_R N_L \right)\;, \\
\Lcal_\Phi :=&\ \Phi \left( \la^Q_{i}\ol{Q}_R {q_L}_i +\la^L_{i}\ol{L}_R {l_L}_i \right) \notag \\
               &\ - \Phi^* \left( \la^U_{i} {\ol{u}_R}_i U_L  + \la^D_{i}{\ol{d}_R}_i D_L
                                        + \la^E_{i} {\ol{e}_R}_i E_L + \la^N_{i} {\ol{\nu}_R}_i N_L
                                            \right) \;.
\label{eq-yukPhi}
\end{align}
Here, we have used $\ti{H}:=i\sigma_2 H^* = (H_-^*, -H_0^{*})$ and $i,j=1,2,3$ run over the three SM generations.

The scalar fields acquire vacuum expectation values (VEVs) given by
$v_{H}:=\vev{H}$, $v_{\phi}:=\vev{\phi}$, and $v_{\Phi}:=\vev{\Phi}$. The charged lepton mass matrix then is given by
\begin{align}
\label{eq-Meg}
\ol{e}^A_R \Mcal^e_{AB} e^B_L
=
\begin{pmatrix}
{\ol{e}_R}_i & \ol{E}_R&  \ol{E}'_R
\end{pmatrix}
\begin{pmatrix}
y^e_{ij} v_H         &   0_i & \la^E_{i} v_\Phi  \\
 0_j                         & \la_e v_H &  \la_V^Ev_\phi \\
 \la^L_j v_\Phi     &  \la_V^L v_\phi & \la'_e v_H
 \end{pmatrix}
\begin{pmatrix}
{e_L}_j  \\ E'_L \\  E_L
\end{pmatrix}\;,
\end{align}
where $A, B = 1,\dots,5$. We define the mass basis via
\begin{align}
 \left[\hat{e}_L\right]_A := \left[\left(U^e_{L} \right)^\dagger \right]_{AB} \left[e_L \right]_B\;,
\quad
 \left[\hat{e}_R\right]_A := \left[\left(U^e_{R} \right)^\dagger \right]_{AB} \left[e_R \right]_B\;,
\end{align}
with unitary matrices that satisfy
\begin{align}
\left( U_R^e \right)^\dagger \Mcal^e U_L^e = \mathrm{diag}\left(m_e, m_\mu, m_\tau, m_{E_1}, m_{E_2} \right)\;.
\end{align}
Here $m_{E_1}$ and $m_{E_2}$ are masses for the extra charged leptons, which are predominantly the VL leptons of the gauge basis.
The mass matrices for the up and down quarks are obtained from $\Mcal^e$ by formally replacing $e\to u, E\to U$, or $e\to d, E\to D$, respectively.

As a consequence of the $\U1'$ charges, only the three standard generations of right-handed (RH) neutrinos have Majorana masses,
\begin{align}
 \Lcal_\mathrm{Maj} = - \frac{1}{2}  \ol{\nu}_{R_i} M^{ij}_\mathrm{Maj} \nu^c_{R_j}\;.
\end{align}
The neutrino Dirac mass matrix is obtained from $\Mcal^e$ by formally replacing $e\to n$ and $E\to N$.
The Majorana masses are assumed to be $\mathcal{O}(10^{14})\,\mathrm{GeV}$, thereby explaining the tiny observed neutrino masses
via a standard type I see-saw mechanism.
As usual, three generations of left-handed (LH) neutrinos have tiny masses of $\mathcal{O}\left(v_H^2 / M_\mathrm{Maj}\right)$
while three generations of RH neutrinos have huge masses $\mathcal{O}(M_\mathrm{Maj})$.
Unlike in the standard type I see-saw case, there are $4$ more degrees of freedom here, which form two Dirac fermions
with masses of $\mathcal{O}(\mathrm{TeV})$.
The mixing of those with the other neutrinos is negligible as it is suppressed by the Majorana mass terms.

The neutral scalar fields are expanded as
\begin{align}
H_0   =  v_H    +\frac{1}{\sqrt{2}}( h + i a_h), \quad
 \Phi =  v_\Phi +\frac{1}{\sqrt{2}}( \chi + i a_\chi), \quad
 \phi =  v_\phi + \sigma.
\end{align}
The CP-odd degrees of freedom, namely  $a_h$ and $a_\chi$,
get eaten by the gauge fields and only the real components $h$ and $\chi$ are physical.
We assume that $\phi$ is a real scalar field.

We parametrize the masses for $\chi$ and $\sigma$ as
\begin{align}
\label{eq-mSdef}
 m_\chi^2 = \lambda_\chi v_\Phi^2\;\quad\mathrm{and}\quad  m_\sigma^2 = \lambda_\sigma v_\phi^2.
\end{align}
Here we have introduced the \textit{effective} quartic couplings $\la_\chi$ and $\la_\sigma$.
The scalar $\chi$ should not be much heavier than the $Z'$ boson,
as long as the effective quartic coupling stays perturbative and the new gauge coupling $g'$ is not tiny.
Importantly, the couplings of $\chi$ are responsible for the mass mixing of SM particles with the VL families. Consequently,
to the extent that this mixing is necessary to fit the muon anomalies, $\chi$ contributes significantly in our fits.
On the other hand, $v_\phi$ could be very large compared with $v_\Phi$ as long as the Yukawa couplings to $\phi$, e.g.\ $\la_V^L$ and $\la_V^E$, are
small enough to prevent the VL fermions from decoupling much above the TeV scale.
The scalar $\sigma$, thus, can be heavy and therefore irrelevant for current observables.
Indeed, contributions from $\sigma$ will be negligible at the best fit points shown below.

\subsection{Yukawa and Gauge Couplings}
The real scalar fields couple to the charged leptons as
 \begin{align}
-  \Lcal_\mathrm{Yukawa} =&\  \frac{1}{\sqrt{2}} h\ \ol{e}_R Y^h_e  e_L
                                  + \frac{1}{\sqrt{2}} \chi\ \ol{e}_R Y^\chi_e  e_L
                                  + \mathrm{h.c.}\;,
 \end{align}
where, in the gauge basis,
\begin{align}
 Y^h_e =
\begin{pmatrix}
 y^e_{ij} & 0_i & 0_i \\
 0_j &   \la_e      & 0  \\
 0_j &    0         & \la'_e
 \end{pmatrix},\quad
 Y^\chi_e =
\begin{pmatrix}
 0_{ij} & 0_i & \la^E_i \\
 0_j   &  0    & 0  \\
 \la^L_j &    0    &  0
 \end{pmatrix},
\end{align}
and completely analogously for the quarks.
The Yukawa coupling matrices in the mass basis are given by
\begin{align}
 \hat{Y}^S_e = \left(U_R^e\right)^\dag Y^S_e U_L^e,\quad\mathrm{where}\quad S = h, \chi\;.
\end{align}

Combining LH and RH fields \`a la $f^A:=(f_L^A,f_R^A)$, the $W$ boson couplings are given by
\begin{align}
\Lcal_W =&\ \frac{g}{\sqrt{2}} W^+_\mu
                   \left[ \ol{u} \gamma^\mu \left( P_{\ol{5}}  P_L + P_5 P_R \right) d
                  +  \ol{\nu}  \gamma^\mu  \left(  P_{\ol{5}}  P_L + P_5 P_R \right) e
                                \right] + \mathrm{h.c.}\;,
\end{align}
where we have used the flavor space projectors
\begin{align}
P_5 := \mathrm{diag}(0,0,0,0,1),\quad\mathrm{and}\quad   \Pfb :=\mathbbm{1}_5-P_5.
\end{align}
The couplings in the mass basis are\footnote{%
Note that here and in the following we neglect effects of $\mathcal{O}\left(v_H^2 / M_\mathrm{Maj}\right)$,
implying that we can treat the left- and right-handed neutrino rotations separately.
}
\begin{align}
\hat{g}^W_{q_L} &= \frac{g}{\sqrt{2}} \left(U_L^u\right)^\dag P_{\ol{5}}\    U_L^d\;, & \quad
\hat{g}^W_{q_R} &= \frac{g}{\sqrt{2}} \left(U_R^u\right)^\dag P_{5}\         U_R^d\;, \\
\hat{g}^W_{\ell_L} &=  \frac{g}{\sqrt{2}} \left(U_L^{n}\right)^\dag P_{\ol{5}}\ U_L^{e}\;,& \quad
\hat{g}^W_{\ell_R} &= \frac{g}{\sqrt{2}} \left(U_R^{n}\right)^\dag P_{5}\      U_R^{e}\;.
\end{align}
Note that there are also right-handed charged current interactions unlike in the SM.

The extended CKM matrix is a $5\times 5$ matrix,
\begin{align}
 \hat{V}_\mathrm{CKM} =  \left(U_L^u\right)^\dag P_{\ol{5}}\ U_L^d.
\end{align}
The $3\times 3$ CKM matrix for the three SM families is not unitary because of the mixing with the VL family.
We remark that also the $5\times 5$ matrix $\hat{V}_\mathrm{CKM}$ is not unitary.

The $Z$ boson couplings are given by\footnote{%
Here we abbreviate the (co)sine of the weak mixing angle as $s_W(c_W)$,
$T^f_3$ and $Q_f$ are respectively the third component of weak isospin and the electromagnetic charge of the fermion $f$,
and five dimensional identity matrices $\mathbbm{1}_5$ in flavor space are implicit where appropriate.}
\begin{align}
 \Lcal_Z =&\ \frac{g}{c_W} Z_\mu  \sum_{f=u,d,e,n} \ol{f}\gamma^\mu
                   \left[  \left( T^f_3 P_{\ol{5}} - Q_f s^2_W \right) P_L
                         + \left( T^f_3 P_{5}       - Q_f s^2_W \right) P_R
                   \right] f.
\end{align}
The couplings for a fermion $f=u,d,e,n$ in the mass basis are given by
\begin{align}
 \hat{g}^Z_{f_{L}} =\frac{g}{c_W} \left[ \left(U_L^f\right)^\dagger \Pfb U_L^f -Q_f s_W^2 \right]\;, \quad
 \hat{g}^Z_{f_{R}} =\frac{g}{c_W} \left[ \left(U_R^f\right)^\dagger P_5 U_R^f -Q_f s_W^2 \right]\;.
\end{align}

Finally, the couplings to the $Z'$ boson are given by
\begin{align}
\Lcal_{Z'} =&\ g' Z'_\mu  \sum_{f=u,d,e,\nu} \ol{f} \gamma^\mu\left(  {Q}'_{f_L} P_L   +  {Q}'_{f_R}  P_R \right) f \notag\\
            =&\ Z'_\mu \sum_{f=u,d,e,n} \ol{\hat{f}}  \gamma^\mu\left( \hg^{Z'}_{f_L} P_L +  \hg^{Z'}_{f_R} P_R \right) \hat{f}\;,
 \end{align}
where the fermions in the mass basis are denoted by $\hat{f}$ and the charge matrices are
\begin{align}
Q'_{{f_L}} = Q'_{{f_R}} = \mathrm{diag}(0,0,0,-1,-1)\;.
\end{align}
In the  mass basis,
\begin{align}
\hat{g}^{Z'}_{f_{L}} =g' \left(U^f_{L} \right)^\dag  Q'_{{f_L}} U_{L}^f\;\quad\mathrm{and}\quad \hat{g}^{Z'}_{f_{R}} =g' \left(U^f_{R} \right)^\dag  Q'_{{f_R}} U_{R}^f\;,
\end{align}
implying that all couplings between $Z'$ and the SM fermions are controlled by the mixing matrices.

Altogether, we find that the model has non-unitary CKM mixing
and tree-level flavor changing neutral currents
mediated by $Z$, $Z'$, the SM Higgs boson, as well as by the new boson $\chi$.
In addition, $W$ bosons also acquire couplings to the right-handed charged currents of SM fermions, which
are constrained by measurements such as neutrino-nucleon scattering~\cite{Bernard:2007cf}.
All of these effects are in general severely constrained by experiments.
However, we find that in our model all those effects are controlled by $\mathcal{O}(m_f^2/M_{\mathrm{VL}}^2)$ coefficients
implying that they are generally suppressed. We prove this analytically in Appendix~\ref{analytical_analysis}.
In agreement with this, both, the unitarity of the $3\times3$ CKM and PMNS matrix
as well as the absence of all
tree-level flavor violating effects for SM fermions,
are restored in the limit of a heavy VL family. That is, the model approaches the SM in the decoupling limit.

\subsection{RGE Evolution of the \texorpdfstring{$\boldsymbol{\U1'}$}{U(1)'} Gauge Coupling}

\begin{figure}[t]
 \centering
\includegraphics[height=75mm]{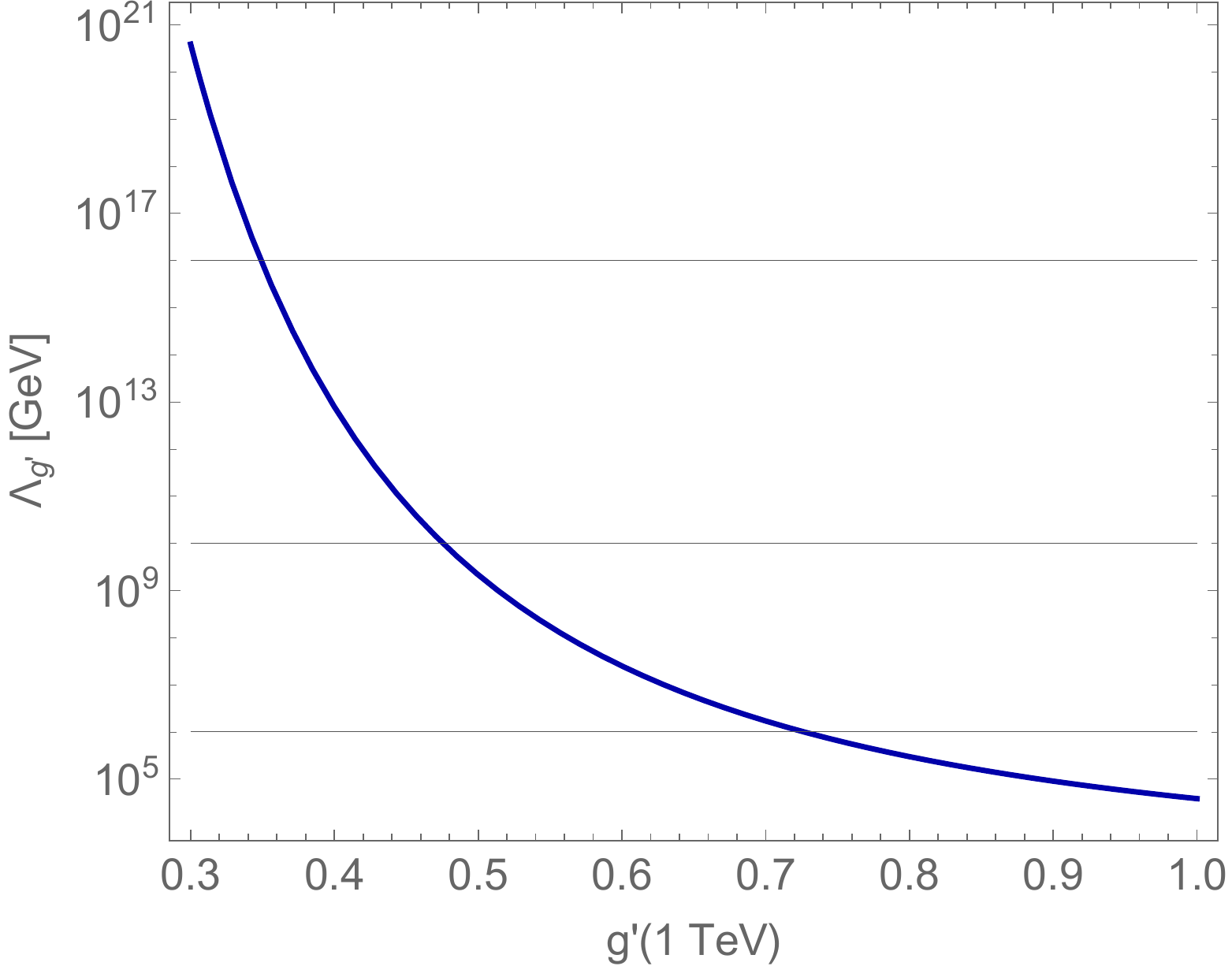}
\caption{Scale of the $\U1'$ Landau pole as a function of $g'$ at $1\,\mathrm{TeV}$}
\label{fig-gpLP}
\end{figure}

The $\U1'$ gauge coupling constant $g'$ should be sufficiently small
at the TeV scale such that it stays perturbative under RGE running up to a scale where UV physics,
such as a Grand Unified Theory (GUT), emerges.
The 1-loop beta function for $g'$ is given by
\begin{align}
 \frac{d g'}{d \ln{\mu}} =  \frac{{g'}^3}{16\pi^2} \frac{65}{3}\;.
\end{align}
This gives rise to a scale of the Landau pole for $g'$,
\begin{align}
 \Lambda_{g'} = \mu_{Z'} \exp\left( \frac{24\,\pi^2}{65 \, {g'(\mu_{Z'})}^2} \right),
\end{align}
where $\mu_{Z'}\sim 1\,\mathrm{TeV}$ is the typical scale of the model.
Figure~\ref{fig-gpLP} shows the scale of the Landau pole in dependence of $g'(\mu_{Z'})$ at the
$\mathrm{TeV}$ scale.
For example, $g'(1\ \mathrm{TeV}) \lesssim 0.35\ (0.48)$ is required for $\Lambda_{g'} \sim 10^{16}\ (10^{10})\,\mathrm{GeV}$.
In our numerical analysis we focus on a situation where the model is correct up to a typical GUT scale of $10^{16}$ GeV,
such that $g'(1\,\mathrm{TeV}) < 0.35$ is required.

\section{Observables}
\label{sec-obs}
In this model, $\Delta a_\mu$ is explained by 1-loop contributions involving the $Z'$ boson and VL leptons.
NP contributions to $C_{9,10}^{(\prime)\mu}$ are provided by tree-level $Z'$ exchange.
NP contributions will also affect observables which are currently consistent with the SM such as
$\Br{\mu}{e\gamma}$, $\Br{\tau}{\mu\gamma}$, $\Br{\tau}{\mu\mu\mu}$, ${B}_q$-$\ol{B}_q$ mixing, \textit{etc.}
The most relevant observables for the muon anomalies will be discussed in the following.
An in-depth discussion of these and further observables is postponed to future work.

\subsection{\texorpdfstring{$\boldsymbol{\Delta a_\mu}$}{amu} and \texorpdfstring{$\boldsymbol{\mathrm{U}(1)'}$}{\U1'} Charge Assignment}
\label{subsec-delamu}

\begin{figure}[t]
 \centering
\includegraphics[height=90mm]{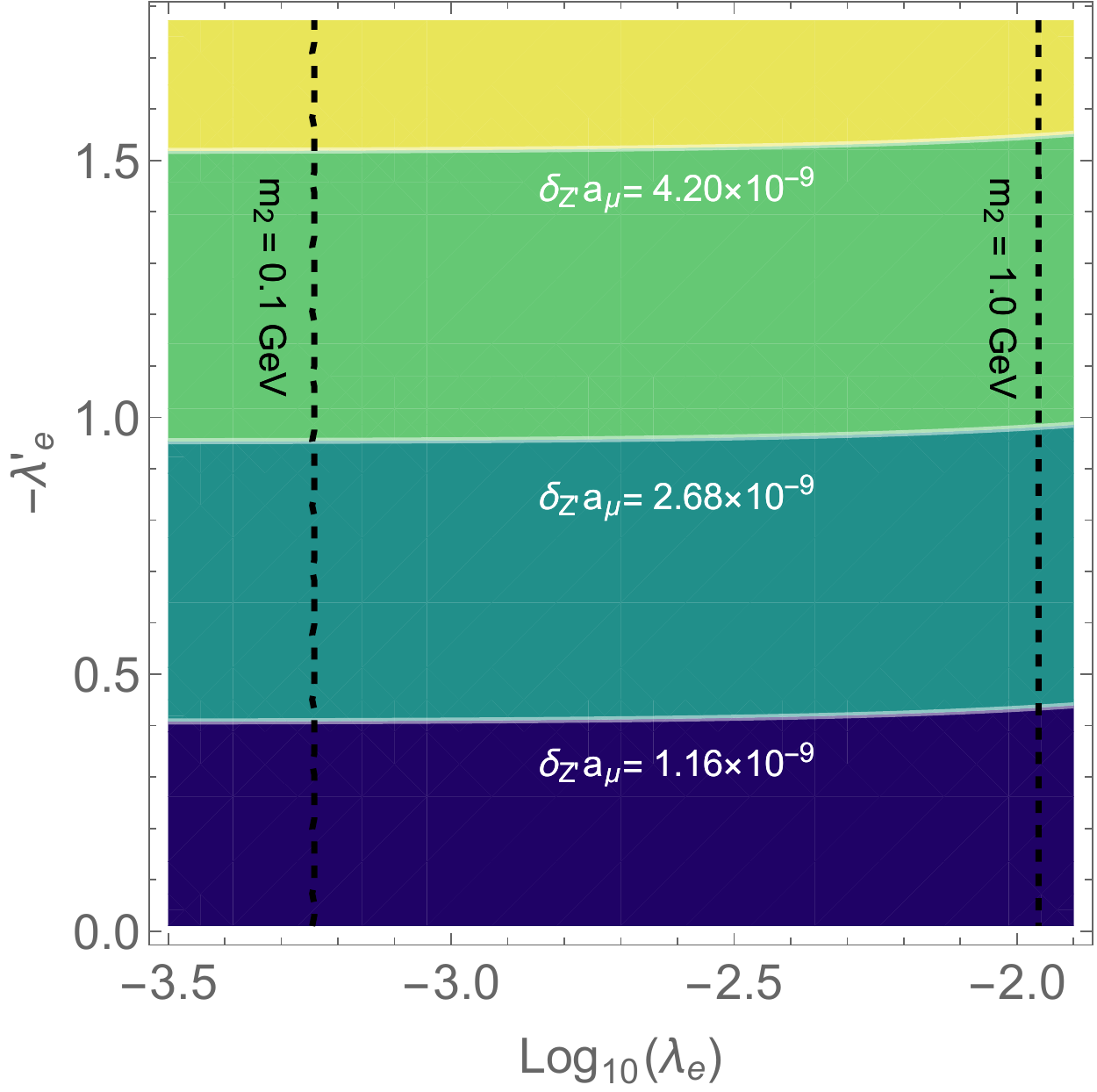}
\caption{The muon mass $m_2$ and $\delta_{Z'} a_\mu$ in dependence of $\la_e$ and $\la'_e$.}

\label{fig-muon}
\end{figure}

The dominant $Z'$ boson contribution to $\Delta a_\mu$ is given by~(see e.g.\ \cite{Jegerlehner:2009ry,Dermisek:2013gta}), 
\begin{align}
\label{eq-delamu}
 \delta_{Z'} a_\mu \simeq -\frac{m_\mu}{8\pi^2m_{Z'}^2} \sum_{a=1,2}
\mathrm{Re} \left( \left[\hg^{Z'}_{\he_L}\right]_{\mu E_a}
\left[{\hg^{Z'}_{\he_R}}\right]^*_{\mu E_a}
         \right)
        m_{E_a} G_{Z}(x_{a})\;,
\end{align}
where $x_{a} := m_{E_a}^2/m_{Z'}^2 $ and the loop function is given by
\begin{align}
\label{eq-loopG}
 G_{Z}(x) := &\ \frac{x^3+3x-6x \ln{(x)}-4}{2(1-x)^3}\;.
\end{align}
The dominant contribution of the scalar $\chi$ is given by
\begin{align}
 \delta_{\chi} a_\mu \simeq& -\frac{m_\mu}{32\pi^2 m_{\chi}^2} \sum_{a = 1,2}
\re{\left[\hat{Y}^\chi_e\right]_{\mu E_a} \left[\hat{Y}^\chi_e\right]_{E_a \mu}}
m_{E_a} G_S(y_{a})
\;,
\end{align}
where $y_{a} := m_{E_a}^2/m_\chi^2 $ and the loop function is
\begin{align}
 G_S(x) := \frac{x^2-4x+2\ln{(x)}+3}{(1-x)^3}\;.
\end{align}
There are also new contributions from loops involving the SM bosons and the VL fermions,
but these are negligible.

Figure~\ref{fig-muon} shows typical values of the muon mass $m_2$ and the $Z'$ contribution to $\delta a_\mu$.
For illustration, $g' = 0.25$, $m_{Z'} = \la^{L,E}_2 v_\Phi = \la^{L,E}_V v_\phi = 500\,\mathrm{GeV}$ and
$y^e_{22} v_H = 0.1\,\mathrm{GeV}$ have been fixed while all other couplings except $\la_e$ and $\la'_e$ are set to zero.
We see that $\abs{\la'_e} \gtrsim 0.4$ and $\la_e \lesssim 10^{-3}$ are required in order to obtain $\Delta a_\mu \gtrsim 10^{-9}$ and $m_\mu \sim 0.1\,\mathrm{GeV}$.
This illustrates how the muon mass is affected by the mixing and enhanced above $m_\mu\sim 0.1$ GeV for $\la_e \gtrsim 10^{-3}$.

We see that the Higgs coupling $\la'_e$ \textit{must} exist in the model to explain $\Delta a_\mu$.
This explains the non-universal charge assignment in Table \ref{tab-extra}:
The $\U1'$ charges of VL-fermions needs to be opposite for $\SU2_{\mathrm{L}}$ doublets and singlets
in order to allow the coupling $\la'_e$. For this reason the $\U1'$ gauge symmetry is incompatible with $\SU5$ unification.
However, it is still compatible with the Pati-Salam gauge group $\SU4\times \SU2_{\mathrm{L}}\times \SU2_{\mathrm{R}}$.

\subsection{\texorpdfstring{$\boldsymbol{\br{\ell_i}{\ell_j \gamma}}$}{BR(li->gamma lj)}}
\label{subsec-LFV}
The branching fraction of $\ell_i\to \ell_j \gamma$ is given by~\cite{Lavoura:2003xp}
\begin{align}
 \br{\ell_i}{\ell_j\gamma} \simeq \frac{1}{\Gamma_{\ell_i}} \frac{\alpha_e\, m^3_{\ell_i}}{1024\pi^4}
\left(\abs{ \sigma_L}^2 + \abs{\sigma_R }^2 \right),
\end{align}
where $m_{\ell_i}$ and $\Gamma_{\ell_i}$ are the mass and total decay width of the lepton $\ell_i$,
while $\alpha_e$ is the electromagnetic fine-structure constant.
The dominant contributions arise from $Z'$ or $\chi$ exchange and they are given by
\begin{align}
\sigma_L \simeq &\   \sum_{a=1,2} \left(\frac{m_{E_a}}{m_{Z'}^2} 
               \left[\hg^{Z'}_R \right]_{j E_a}\left[\hg^{Z'}_L \right]_{E_a i} G_Z(x_a)
             + \frac{m_{E_a}}{4m_{\chi}^2}  
                     \left[\hY^{\chi}_e \right]_{j E_a}\left[\hY^{\chi}_e \right]_{E_ai} G_S(y_a)
                \right),
\end{align}
and $\sigma_R$ which is given by formally replacing $L\to R$ and $\hY^{\chi}_e \to \left(\hY^\chi_e \right)^\dagger $
in the above expression.
Other contributions, involving the SM bosons or $\sigma$, only amount to sub-percent corrections at our 
best fit points.

\subsection{Wilson Coefficients for \texorpdfstring{$\boldsymbol{\bsll}$}{bsll}}
The Wilson coefficients defined in Eqs.~(\ref{eq-defC9C10}) are given by
 \begin{align}
 C^{\ell}_9 =&\  -\frac{\sqrt{2}}{4 G_F} \frac{4\pi}{\alpha_e}\frac{1}{V_{tb}V_{ts}^*} \frac{1}{2m^2_{Z'}}
                \left[\hg^{Z'}_{d_L} \right]_{23} \left[\hg^{Z'}_{e_{R}}+\hg^{Z'}_{e_L}  \right]_{ii}, \\
 C^{\ell}_{10} =&\ -\frac{\sqrt{2}}{4 G_F} \frac{4\pi}{\alpha_e}\frac{1}{V_{tb}V_{ts}^*} \frac{1}{2m^2_{Z'}}
                \left[\hg^{Z'}_{d_L} \right]_{23} \left[\hg^{Z'}_{e_{R}}-\hg^{Z'}_{e_L}  \right]_{ii}, \\
 C^{\prime\ell}_9 =&\  -\frac{\sqrt{2}}{4 G_F} \frac{4\pi}{\alpha_e}\frac{1}{V_{tb}V_{ts}^*} \frac{1}{2m^2_{Z'}}
                \left[\hg^{Z'}_{d_R} \right]_{23} \left[\hg^{Z'}_{e_{R}}+\hg^{Z'}_{e_L}  \right]_{ii}, \\
 C^{\prime\ell}_{10} =&\ -\frac{\sqrt{2}}{4 G_F} \frac{4\pi}{\alpha_e} \frac{1}{V_{tb}V_{ts}^*} \frac{1}{2m^2_{Z'}}
                \left[\hg^{Z'}_{d_R} \right]_{23} \left[\hg^{Z'}_{e_{R}}-\hg^{Z'}_{e_L}  \right]_{ii},
\end{align}
where $i = 1,2,3$ for $\ell = e,\mu,\tau$, respectively.

We refer to the recent two-dimensional analysis of Ref.~\cite{Aebischer:2019mlg} and adopt the best fit values of the Wilson coefficients as
\begin{align}
 &(\mathrm{I})& &\mathrm{Re}\, C_9^\mu = -0.7 \pm 0.3,& &\mathrm{Re}\,C^\mu_{10} = 0.4 \pm 0.25\;;& \\\label{eq:BFWCII}
 &(\mathrm{II})& &\mathrm{Re}\,C_9^\mu = -1.04 \pm 0.24,& &\mathrm{Re}\,C^{\prime\mu}_{9} = 0.48 \pm 0.30\;.&
\end{align}
Note that the $Z'$ should not introduce sizable Wilson coefficients for the electron
because that would generically also induce sizable violation of lepton flavor in $\mu \to e \gamma$.
Although it has been pointed out that some flavor universal contributions seem to be favored \cite{Aebischer:2019mlg,Alguero:2019ptt,Datta:2019zca},
we do not discuss this possibility in the present paper.

\subsection{Neutral Meson Mixing}
\begin{table}[t]
\centering
\caption{\label{tab-valOs}
Numerical values of the operators $O^a_i :=\bra{\ol{M}} Q^a_i \ket{M}/(2 m_{M})$ 
at $\mu_B = 1\,\mathrm{TeV}$.
The corresponding right-right (RR) operators have the same values as the LL operators.}
\begin{tabular}{c|ccccc} \hline
                         & $O^{\text{VLL}}_1(\mu_B)$ &$O^{\text{LR}}_1(\mu_B)$ &$O^{\text{LR}}_2(\mu_B)$ & $O^{\text{SLL}}_1(\mu_B)$ &  $O^{\text{SLL}}_2(\mu_B)$\\ \hline\hline
$K\text{-}\ol{K}$         & $0.00159$ & $-0.159$   & $0.261$ & $-0.0761$ & $-0.132$ \\
$B_d\text{-}\ol{B}_d$     & $0.0465$  & $-0.186$   & $0.241$ & $-0.0909$ & $-0.167$ \\
$B_s\text{-}\ol{B}_s$     & $0.0701$  & $-0.264$   & $0.338$ & $-0.136$  & $-0.252$ \\
$D\text{-}\ol{D}$         & $0.0162$  & $-0.157$   & $0.227$ & $-0.0845$ & $-0.152$ \\
\hline
\end{tabular}
\end{table}

There are strong constraints on neutral meson mixing~\cite{Buras:2012jb,DiLuzio:2017fdq}.
The relevant effective Hamiltonian is given by
\begin{align}
\Hcal^{\Delta F=2}_\mathrm{eff}= \sum_{i,a} C_i^a  Q_i^a,
\end{align}
where $(i,a) = (1,\mathrm{VLL}), (1,\mathrm{VRR}), (1,\mathrm{LR}), (2,\mathrm{LR}), (1,\mathrm{SLL}), (2,\mathrm{SLL}), (1,\mathrm{SRR}), (2,\mathrm{SRR})$.
The four-fermi operators are defined as
\begin{align}
 Q_1^\mathrm{VLL} =&\ \left(\ol{F}^\alpha \gamma_\mu P_L f^\alpha \right)\left(\ol{F}^\beta \gamma^\mu P_L f^\beta \right)\;,&
 Q_1^\mathrm{VRR} =&\ \left(\ol{F}^\alpha \gamma_\mu P_R f^\alpha \right)\left(\ol{F}^\beta \gamma^\mu P_R f^\beta \right)\;,& \\
 Q_1^\mathrm{LR\phantom{R}}  =&\ \left(\ol{F}^\alpha \gamma_\mu P_L f^\alpha\right) \left(\ol{F}^\beta \gamma^\mu P_R f^\beta \right)\;,&
 Q_2^\mathrm{LR\phantom{R}}  =&\ \left(\ol{F}^\alpha P_L f^\alpha \right) \left(\ol{F}^\beta  P_R f^\beta \right)\;,& \\
 Q_1^\mathrm{SLL} \,=&\ \left(\ol{F}^\alpha P_L f^\alpha \right)\left(\ol{F}^\beta  P_L f^\beta \right)\;,&
 Q_2^\mathrm{SLL} \,=&\ \left(\ol{F}^\alpha \sigma_{\mu\nu} P_L f^\alpha \right)\left(\ol{F}^\beta \sigma^{\mu\nu} P_L f^\beta \right)\;,&  \\
 Q_1^\mathrm{SRR} =&\ \left(\ol{F}^\alpha P_R f^\alpha \right)\left(\ol{F}^\beta  P_R f^\beta \right)\;,&
 Q_2^\mathrm{SRR} =&\ \left(\ol{F}^\alpha \sigma_{\mu\nu} P_R f^\alpha \right)\left(\ol{F}^\beta \sigma^{\mu\nu} P_R f^\beta \right)\;.&
\end{align}
Here $\alpha$ and $\beta$ are color indices and $(F, f) = (b,d), (b,s), (s,d)\,\text{or}~(c, u)$ for
$B_d$-$\ol{B}_d$, $B_s$-$\ol{B}_s$, $K$-$\ol{K}$ or $D$-$\ol{D}$ mixing, respectively.
We focus here on $B_q$-$\ol{B}_q$ ($q=d,s$) mixing since these are the most relevant for the $\bsll$ anomalies.

The Wilson coefficients induced by $Z'$ or neutral scalar exchange, including $\mathcal{O}(\alpha_s)$ QCD corrections are given by~\cite{Buras:2012fs}
\begin{align}
C_1^\mathrm{VLL}(\mu)=&\ \left[ 1+\frac{\alpha_s}{4\pi}\left(-2\log\frac{m_{Z'}^2}{\mu^2} + \frac{11}{3}\right) \right]
                                \frac{g^{Z'}_L g^{Z'}_L} {2\,m_{Z'}^2},  \\
C_1^\mathrm{LR}(\mu)= &\ \left[1+\frac{\alpha_s}{4\pi}\left(-\log\frac{m_{Z'}^2}{\mu^2} - \frac{1}{6}\right) \right]
                                  \frac{g^{Z'}_L g^{Z'}_R} {m_{Z'}^2}
                                  - \left( -\frac{3}{2}\frac{\alpha_s}{4\pi} \right)
                                 \frac{y^{\chi}_L y^{\chi}_R} {2\,m_{\chi}^2},  \\
C_2^\mathrm{LR}(\mu)= &\ \frac{\alpha_s}{4\pi}\left(-6\log\frac{m_{Z'}^2}{\mu^2} - 1 \right)
                              \frac{g^{Z'}_L g^{Z'}_R} {m_{Z'}^2}
                              - \left(1-\frac{\alpha_s}{4\pi}\right)
                                 \frac{y^{\chi}_L y^{\chi}_R} {2\,m_{\chi}^2},  \\
C_1^\mathrm{SLL}(\mu) =&\ - 
                     \left[1+\frac{\alpha_s}{4\pi}\left(-3\log\frac{m_\chi^2}{\mu^2} + \frac{9}{2}\right)\right]
                                        \frac{y^{\chi}_L y^{\chi}_L} {4\,m_{\chi}^2} ,   \\
C_2^\mathrm{SLL}(\mu)=&\  -
                                  \frac{\alpha_s}{4\pi}\left(-\frac{1}{12}\log\frac{m_\chi^2}{\mu^2}+\frac{1}{8}\right)
                                      \frac{y^{\chi}_L y^{\chi}_L} {4\,m_{\chi}^2} , 
\end{align}
 where
\begin{align}
g_L^{Z'} =  \left[\hg^{Z'}_{d_L}\right]_{3k},\  g_R^{Z'} = \left[\hg^{Z'}_{d_R}\right]_{3k},\
y_L^\chi  = \left[\hY^\chi_{d} \right]_{3k},\ y_R^\chi  = \left[\hY^\chi_{d} \right]^*_{k3},\
\end{align}
and $k=1(2)$ for $B_{d(s)}$-$\ol{B}_{d(s)}$ mixing, respectively.
The analogous Wilson coefficients $C_1^\mathrm{VRR}$ and $C_{1,2}^\mathrm{SRR}$ are obtained by formally replacing $L\to R$.
The off-diagonal element of the $B_q(q=d,s)$ meson mass matrix is given by
\begin{align}
 M_{12}^*(B_q) = M^{\mathrm{SM}*}_{12}(B_q) + \frac{1}{2\,m_{B_q}} \sum_{i,a} C_i^a(\mu)\,\bra{\ol{B}_q} Q^a_i(\mu)\ket{B_q}\;,
\end{align}
where the first term is the SM contribution and $m_{B_q}$ is the meson's physical mass.
The SM contribution for $B_q$-$\ol{B}_q$ mixing is given by
\begin{align}
\label{eq-M12SM}
M_{12}^{\mathrm{SM}*}(B_q) =
           \frac{G_F^2}{12\pi^2}  m_W^2 \,(\lambda^{(q)}_t)^2 \, S_0(x_t) \,\eta_B\, m_{B_q}\, f^2_{B_q}\,  \hat{B}_{B_q}\;.
\end{align}
Here, $\la_t^{(q)} = V_{tb}^* V_{tq}$, $S_0(x_t=m_t^2/M_W^2)\approx 2.32$ is the Inami-Lim loop function~\cite{Inami:1980fz},
$\eta_B=0.55\pm0.01$ \cite{Buras:1990fn,Urban:1997gw} quantifies the short distance radiative corrections, while
$f_{B_q}$ and  $\hat{B}_{B_q}$ denote the corresponding decay constant and SM hadronic matrix element.
The SM and necessary BSM hadronic matrix elements are calculated by lattice collaborations,
and their values at $1\,\mathrm{TeV}$ according to our own evaluation are listed in Table~\ref{tab-valOs}.
Values for $K$-$\ol{K}$ and $D$-$\ol{D}$ mixing are also listed for completeness.
All hadronic matrix elements for Kaon oscillations and the value of
 $f^2_{B_q} \hat{B}_{B_q}$
have been taken from Ref.~\cite{Aoki:2019cca}, while those for $f_{B_q}^2 B^{(2\text{-}5)}_{B_q}$ and $B_D^{(1\text{-}5)}$
are taken from Refs.~\cite{Carrasco:2013zta,Bazavov:2016nty} and Ref.~\cite{Carrasco:2014uya}, respectively.
The QCD running between the respective lattice scales and $\mu = 1\,\mathrm{TeV}$
has been calculated based on the anomalous dimensions shown in Ref.~\cite{Buras:2001ra}.

The observables for $B_q$-$\ol{B}_q$ mixing are defined as
 \begin{align}
\label{eq-defmm}
 \Delta M_d :=&\ 2 \abs{M_{12}(B_d)}\;,&
S_{\psi K_s} :=&\ \sin\left(\mathrm{Arg}\left[M_{12}(B_d)\right]\right)\;,& \\
 \Delta M_s :=&\ 2 \abs{M_{12}(B_s)}\;,&
S_{\psi \phi}\,\, :=&\ - \sin\left(\mathrm{Arg}\left[M_{12}(B_s)\right]\right)\;.&
 \end{align}
The mass differences $\Delta M_d$ and $\Delta M_s$ are measured with high accuracy
and theoretical uncertainties are prevailing.
The dominant theoretical uncertainties arise from the CKM elements, hadronic matrix elements, and NLO QCD corrections.
Altogether we find $15.6\% \ (14.1\%)$ relative uncertainty for $\Delta M_d\ (\Delta M_s$).
Note that unlike the analyses in e.g.\ Refs~\cite{DiLuzio:2017fdq,King:2019lal}, 
we cannot reduce the uncertainties by assuming exact unitarity of the CKM matrix here,
simply because CKM unitarity is not guaranteed in our model.
We therefore have to rely on the measured CKM matrix elements and their respective errors.
Despite the possible CKM non-unitarity, we still use formulas for the SM contributions which are obtained
under the implicit assumption of exact CKM unitarity (i.e.\ a working GIM mechanism).
This adds some additional theoretical uncertainty which is hard to quantify.
As the CKM matrix at our fit points is still approximately unitary to the observed degree we neglect this additional uncertainty.

We find that the pattern (II) of the Wilson coefficients for $\bsll$ (cf.\ Eq.~\eqref{eq:BFWCII}) is disfavored,
because this would cause large $Z'$ contributions to $\Delta M_s$.
A large negative contribution $\mathrm{Re}\,C_9^\mu$
together with a positive $\mathrm{Re}\,C_9^{\prime\mu}$
requires a relative sign difference between
$\mathrm{Re}\,\left[g^{Z'}_{d_L} \right]_{23}$ and
$\mathrm{Re} \left[g^{Z'}_{d_R} \right]_{23}$.
Since the hadronic matrix element $O^{\text{LR}}_1$ is sizable and negative,
this would imply a large and positive left-right contribution to $\Delta M_{s}$.
However, as the current SM prediction is already larger than the experimental value,
therefore the $Z'$ coupling with $\mathrm{Re}\,C_9^{\prime\mu}$ is strongly disfavored.
For this reason we could not find any good fit points for the pattern (II).

\subsection{Neutrino Trident Production}
The so-called neutrino trident production $\nu_\mu \to \nu_\mu \mu\mu$ off a nucleus
is a rare process that has been observed at a rate consistent with SM expectations \cite{Geiregat:1990gz,Mishra:1991bv,Adams:1999mn}.
This process can also be mediated by $Z'$ exchange and therefore constitutes an important bound on NP scenarios~\cite{Altmannshofer:2014cfa,Altmannshofer:2014pba,
Magill:2016hgc,Ge:2017poy,Ballett:2018uuc,Altmannshofer:2019zhy}.
The ratio of the cross section including NP at the CCFR experiment can be estimated as~\cite{Altmannshofer:2019zhy}
\begin{align}
 \frac{\sigma_\mathrm{CCFR}}{\sigma^\mathrm{SM}_\mathrm{CCFR}}
\simeq \frac{(1+4s_W^2 + \Delta g^V_{\mu\mu\mu\mu})^2 +
1.13 (1-\Delta g^A_{\mu\mu\mu\mu})^2}{(1+4s_W^2 )^2 + 1.13 }\;,
\end{align}
with a current experimental limit of $\sigma/\sigma^{\mathrm{SM}}=0.82\pm0.28$ at $95\%\,\mathrm{C.L.}$. 
The effective four-fermi couplings $\Delta g^{V,A}_{\mu\mu\mu\mu}$ in our model are given by
\begin{align}
 \Delta g^{V,A}_{\mu\mu\mu\mu} = \frac{1}{\sqrt{2}G_F m_{Z'}^2 }
                                                    \left[ g^{Z'}_{\nu} \right]_{\nu_\mu\nu_\mu}
                       \left(\left[ g^{Z'}_{e_R} \right]_{22}\pm\left[ g^{Z'}_{e_L} \right]_{22} \right)\;,
\end{align}
where $\left[ g^{Z'}_{\nu} \right]_{\nu_\mu\nu_\mu}$ is defined in the flavor basis,
\begin{align}
 g^{Z'}_{\nu} = g' U_{e_L}^\dagger 
Q'_{n_L}  U_{e_L}.
\end{align}
This constraint is particularly relevant for light $Z'$'s and quickly becomes insensitive
to NP once the $Z'$ is heavier than a few $100\,\mathrm{GeV}$.

\subsection{Gauge Kinetic Mixing}
A potentially light $Z'$ boson can experience sizable gauge kinetic mixing
with the $\U1_{\mathrm{Y}}$ gauge boson, namely the $Z$ of the SM.
The $Z$-$Z'$ mixing parameter $\eps$ is estimated as
\begin{align}
 \eps \simeq \frac{g_Y g'}{6\pi^2} \log{ \left(\frac{m_E^2}{m_L^2}\,\frac{m_Q^2 m_D^2}{m_U^4}\right) }\;,
\end{align}
where $m_F$ $(F=L,E,Q,U,D)$ are the VL mass terms for the VL fermions
and $g_Y$ is the $\U1_{\mathrm{Y}}$ gauge coupling constant.
Current experimental limits are summarized in Ref.~\cite{Hook:2010tw}.
Values of $\eps\sim 0.05$ cannot be ruled out if the $Z'$
is heavier than a few $100\,\mathrm{GeV}$.

\section{Results}
\label{sec-rslt}
\subsection{\texorpdfstring{$\boldsymbol{\chi^2}$}{chi2} Fitting}

We minimize the $\chi^2$ function
\begin{align}
 \chi^2(x) := \sum_{I} \frac{(y_I(x) - y_I^0)^2}{\sigma_I^2},
\end{align}
where $x$ is a point in the parameter space, while
$y_I(x)$ is the value of observable $I$ with central value $y_I^0$ and uncertainty $\sigma_I$.
Observables we fit to include the SM fermion masses, CKM matrix element absolute values and relative phases,
SM particle branching fractions (including flavor violating decays),
neutral meson mixing, $\Delta a_\mu$, $C_{9,10}^{(\prime)l}$, and some others.
In total we consider $98$ observables and they are all listed in appendices~\ref{app:BFA} and \ref{app:BFB}.

In total there are $65$ input parameters represented by $x$ in our analysis.
The bosonic sector has $5$ parameters,
\begin{align}
 m_{Z'},\ v_\phi,\ g',\ \lambda_\chi,\ \lambda_\sigma,
\end{align}
which are the $Z'$ mass, the VEV of $\phi$, the $\U1'$ gauge coupling constant,
and the effective quartic couplings defined in Eq.~(\ref{eq-mSdef}), respectively.
All other parameters are Yukawa couplings as defined in Eqs.~(\ref{eq-yukSM})-(\ref{eq-yukPhi}).
For the neutrino sector we only consider the couplings $\lambda_n^{(\prime)}$ and $\lambda_V^{N}$
which are relevant for the new $\mathrm{TeV}$-scale state.
The remaining Yukawa couplings $y^n_{ij}$, $\lambda^N_{j}$
as well as the heavy Majorana masses are not varied in our analysis,
because these are relevant only for the details of masses and mixings of the SM neutrinos
(and the RH Majorana neutrinos). Thus, effects suppressed by the Majorana masses are neglected.
We expect that the number of parameters in the neutrino sector is sufficient to explain
the observed neutrino mass squared differences and the PMNS matrix without
changing any observables studied in our analysis.
We assume that all Yukawa couplings are real except for $y^{u,d}_{13}$ and $y^{u,d}_{31}$.
Altogether there are then $60$ real parameters for the Yukawa couplings.
All Yukawa couplings and effective quartic coupling
values are restricted to be smaller than unity.
Furthermore, as already discussed at the end of Section~\ref{sec-model}, $g' < 0.35$ is required
so that the gauge coupling $g'$ stays perturbative up to $\sim 10^{16}\,\mathrm{GeV}$.

\subsection{Best Fit Points}
We find two best fit points A and B with $\chi^2 = 25.1$ and $\chi^2 =24.9$, respectively, for $98-65=33$ d.o.f.\ .
The values of all observables and the corresponding input parameters are listed in appendices~\ref{app:BFA} and \ref{app:BFB}.
The values of selected observables are shown in Table~\ref{tab-obs}.
Masses and dominant decay modes of new particles are summarized in Tables~\ref{tab-massdecayA} and~\ref{tab-massdecayB}.
\begin{table}[t]
 \centering
\caption{\label{tab-obs}
Values of selected observables at the best fit points A and B.}
\begin{tabular}[t]{c|ccc}\hline
 Parameters                         & Point A            & Point B              & Remark                              \\   \hline\hline
$\chi^2$                            & $25.1$               & $24.9$                 & $N_\mathrm{inp}= 65$, $N_\mathrm{obs}=98$  \\
$g'$                                & $0.266$              & $0.306$                & $\lesssim 0.35$ for $\Lambda_{g'} =10^{16}\,\mathrm{GeV}$           \\
$(v_\Phi,\ v_\phi)\,[\mathrm{TeV}]$           & $(1.31,\ 4.36)$   & $(0.872,\ 2.92)$       & \\ \hline
 Observables                        & Point A            & Point B              & Data                              \\   \hline\hline
 $\Delta a_\mu \times 10^9$         & $2.56$               & $2.43$                 & $2.68(76)$~\cite{Tanabashi:2018oca}      \\
 $\br{\mu}{e\gamma}\times 10^{13}$  & $3.58$               & $2.10$                 & $<4.2\,(90\%\mathrm{C.L.})$~\cite{Tanabashi:2018oca}             \\
 $\br{\tau}{\mu\gamma}\times10^{8}$ &$1.96\times 10^{-5}$  &  $4.91\times10^{-2}$   &   $<4.4\,(90\%\mathrm{C.L.})$~\cite{Tanabashi:2018oca}   \\
 $\br{\tau}{\mu\mu\mu}\times10^{8}$ &$3.03\times 10^{-5}$  & $7.42\times 10^{-3}$   &   $<2.1\,(90\%\mathrm{C.L.})$~\cite{Tanabashi:2018oca}  \\
\hline
 $\mathrm{Re}\,C_9^\mu$             & $-0.725$               & $-0.571$               & $-0.7(3)$~\cite{Aebischer:2019mlg}  \\
 $\mathrm{Re}\,C_{10}^\mu$          & $0.320$                & $0.316$                & $0.4(2)$~\cite{Aebischer:2019mlg} \\
 $\Delta M_d\,[\mathrm{ps}^{-1}]$   & $0.612$               & $0.599$               & $0.507(81)$~\cite{Tanabashi:2018oca}  \\
 $\Delta M_s\,[\mathrm{ps}^{-1}]$   & $19.4$                & $19.8$                & $17.8(2.5)$~\cite{Tanabashi:2018oca}  \\
 $S_{\psi K_s}$                     & $0.688$               & $0.686$               & $0.695(19)$~\cite{Amhis:2016xyh}  \\
$S_{\psi \phi}$                     & $0.0374$               & $0.0363$               & $0.021(31)$~\cite{Amhis:2016xyh}  \\
\hline\hline
\end{tabular}
\end{table}
\begin{table}[t]
\centering
\caption{\label{tab-massdecayA}
Masses, total widths, and branching fractions (Br) at point A.
Decay~1(2) denote the (next to) dominant decay mode.}
 \begin{tabular}[t]{c|cc|cccc}\hline
                  & Mass [GeV] & Width [GeV] & Decay~1 & Br & Decay~2 & Br \\ \hline\hline
 \text{$Z'$} & 494.7 & 0.7723  & \text{$\mu\mu $} & 0.4058 & \text{$\nu \nu$} & 0.3529 \\
 \text{$\chi$} & 1314. & 2.290 & \text{$N_1 N_1$} & 0.2934 & \text{$E_1 \mu$}      & 0.1477 \\
 \text{$\sigma$} & 4345. & 3.667 & \text{$U_1 t$} & 0.2516 & \text{$D_1 b$} & 0.2476 \\ \hline
 \text{$E_1$} & 267.2 & $2.\times10^{-6}$ & \text{$h\mu $} & 0.6774 & \text{$Z\mu $} & 0.1693 \\
 \text{$N_1$} & 359.2 & 0.5505 & \text{$WE_1$} & 1.000 & \text{$W\mu $} & 0.0000 \\
 \text{$E_2$} & 442.4 & 1.783   & \text{$ZE_1$} & 0.8646 & \text{$WN_1$} & 0.1084 \\
 \text{$N_2$} & 4357. & 0.0019 & \text{$W\mu $} & 0.3745 & \text{$Z \nu$} & 0.1872 \\ \hline
 \text{$D_1$} & 2120. & 1.535 & \text{$Z'b$} & 0.4514 & \text{$W t$} & 0.3629 \\
 \text{$U_1$} & 2120. & 1.538 & \text{$Z't$} & 0.4552 & \text{$ht$} & 0.1840 \\
 \text{$D_2$} & 2947. & 1.029 & \text{$WU_1$}& 0.4983 & \text{$ZD_1$} & 0.2493 \\
 \text{$U_2$} & 4252. & 1.042 & \text{$WD_1$} & 0.4901 & \text{$ZU_1$} & 0.2450 \\
\hline
\end{tabular}
\end{table}
\begin{table}[t]
\centering
\caption{\label{tab-massdecayB}
Masses, total widths, and branching fractions (Br) at point B.
Decay~1(2) denote the (next to) dominant decay mode.}
 \begin{tabular}[t]{c|cc|cccc}\hline
               & Mass [$\mathrm{GeV}$] & Width [$\mathrm{GeV}$] & Decay~1 & Br & Decay~2 & Br \\ \hline\hline
 \text{$Z'$}          & 377.1 & 0.1112 & \text{$\mu \mu $} & 0.5193 & \text{$\nu \nu$} & 0.4793 \\
 \text{$\chi$}      & 135.4 & \text{9.$\times 10^{-9}$}     & \text{$\mu\mu $} & 0.8094 & \text{$b s$} & 0.0945 \\
 \text{$\sigma$} & 2915. & 5.575 & \text{$E_2 E_2$} & 0.2688 & \text{$E_1 E_1$} & 0.1486 \\ \hline
 \text{$N_1$} & 280.2 & 0.0003 & \text{ $W \mu $} & 0.5232     & \text{$Z \nu$}   & 0.2578 \\
 \text{$E_1$} & 571.9 & 0.4464 & \text{$\chi \mu $} & 0.5495   & \text{$Z' \mu $} & 0.3684 \\
 \text{$N_2$} & 580.7 & 0.4718 & \text{$\chi \nu$}  & 0.5427    & \text{$Z' \nu$  } & 0.3741 \\
 \text{$E_2$} & 1174. & 23.07   & \text{$WN_2$} & 0.4698 & \text{$Z E_1$} & 0.2340 \\ \hline
 \text{$D_1$} & 1548. & 0.3587 & \text{$Z's$} & 0.2889 & \text{$\chi s$} & 0.2874 \\
 \text{$U_1$} & 1548. & 0.3594 & \text{$Z'c$} & 0.2833 & \text{$\chi c$} & 0.2818 \\
 \text{$D_2$} & 2902. & 0.1854 & \text{$Z'b$} & 0.4448 & \text{$\chi b$} & 0.4432 \\
 \text{$U_2$} & 2915. & 0.0441 & \text{$W D_1$} & 0.3897 & \text{$h U_1$} & 0.1969 \\
\hline\hline
 \end{tabular}
\end{table}

At both best fit points, deviations from the SM in $\Delta a_\mu$ and $C^\mu_9$, $C^\mu_{10}$ are explained by NP contributions.
Besides a dominant positive loop contribution to $\Delta a_\mu$ involving the $Z'$,
there is a slight cancellation from a negative contribution of the scalar $\chi$ at both points.
All observables are fit within their $2\sigma$ ranges at both points,
except for $\Delta a_e$ which cannot be explained in the SM either, see e.g.\ \cite{Davoudiasl:2018fbb,Crivellin:2018qmi,Liu:2018xkx,Dutta:2018fge,Parker:2018vye,Han:2018znu,Endo:2019bcj}.
We could not find points which have $\Delta a_e$ near the experimental value while LFV processes, especially $\br{\mu}{e\gamma}$, are sufficiently suppressed.
The only observables with pulls exceeding $1\sigma$ are $\br{\mu}{e\gamma}$, $\Delta M_{B_d}$, and some absolute values of CKM elements.
Interestingly, the NP correction to $B_{d}\text{-}\ol{B}_d$ from $Z'$-boson exchange is sizable,
while $B_{s}\text{-}\ol{B}_s$ mixing is within its $1\sigma$ range (taking into account the experimental uncertainties for $\abs{V_{ts}}$ and $\abs{V_{tb}}$).

The $3\times 3$ CKM matrix of the SM families, 
$\left[V_\mathrm{CKM} \right]_{ij} := [\hat{V}_\mathrm{CKM}]_{ij}$,  
is almost unitary with deviations smaller than
\begin{align}
 \left|  V_{\mathrm{CKM}}^\dagger V_{\mathrm{CKM}} - \mathbbm{1}_3 \right| \lesssim
5.1 ~(1.9) \times 10^{-8},  
\end{align}
at the point A(B).
The full $5\times 5$ CKM matrices at the best fit points are shown 
in appendices~\ref{app:BFA} and \ref{app:BFB}.

\subsection{Phenomenology}
\label{subsec-pheno}
We now discuss the phenomenology of this model at the best fit points.
Since the $Z'$ gauge boson and the VL leptons are comparatively light,
constraints from direct searches at the LHC and from muon flavor physics
are both very important.

\subsubsection{\texorpdfstring{$\boldsymbol{Z'}$}{Z'} Physics}
The $Z'$ boson mass is $494.7(377.1)\,\mathrm{GeV}$ at the best fit point A(B).
There are strong limits on such a comparatively light $Z'$ from direct searches at the LHC.
Other important constraints on the $Z'$ mass arise from
the so-called neutrino trident production as well as from gauge kinetic mixing with the $Z$ boson.
However, we will see that despite its relative lightness, the $Z'$ boson is still sufficiently heavy to evade these bounds.

General LHC limits on $Z'$ bosons responsible for $\bsll$ anomalies are studied in Refs.~\cite{Kohda:2018xbc,Allanach:2019mfl}.
The most stringent bound from the LHC on our model comes from resonance searches in the dimuon channel,
\begin{align}
  p p \to Z' \to \mu^+\mu^-\;.
 \end{align}
This is particularly pronounced in our model,
as the dominant decay modes of the $Z'$ are $Z'\to \mu^+\mu^-$ and $Z'\to \nu\nu$,
as shown in Tables~\ref{tab-massdecayA} and \ref{tab-massdecayB}.
Exclusion bounds are given in Ref.~\cite{Aad:2019fac} based on $139\,\mathrm{fb}^{-1}$ of data.
We have calculated the fiducial cross section, as defined in Ref.~\cite{Aad:2019fac}, using \texttt{MadGraph5$\_$2$\_$6$\_$5}~\cite{Alwall:2014hca} 
based on an \texttt{UFO}~\cite{Degrande:2011ua} model file generated with \texttt{FeynRules$\_$2$\_$3$\_$32}~\cite{Christensen:2008py,Alwall:2014hca}.
The fiducial cross section is $0.477(0.482)\,\mathrm{fb}$ at the point A(B).
This is roughly 4 (6) times smaller than the experimental limit for the respective $Z'$ masses~\cite{Aad:2019fac}.
The production cross sections are very suppressed because the $Z'$ couplings to the SM quarks are at most $\mathcal{O}\left(10^{-3}\right)$.
We stress that both of our fit points realize solutions to the observed anomalies where
the dimuon coupling of the $Z'$ is maximal, while the $Z'bs$ coupling is minimal.

The ratio of neutrino trident production at CCFR is $1.014 ~(1.006)$,
very close to the SM and well in agreement with the experimental constraints.
The gauge kinetic $Z-Z'$ mixing at the best fit points is estimated
as $\eps\sim 2.0\times 10^{-3}\ (7.6\times 10^{-4})$
and therefore also well below the experimental bounds.

\subsubsection{Lepton Flavor Physics}
In general, a variety of charged LFV processes are present in this model.
As discussed in Subsection~\ref{subsec-LFV}, loops involving $Z'$ or the scalar $\chi$ together with VL leptons
can lead to chirally enhanced contributions to $\ell_1 \to \ell_2 \gamma$ processes.
Furthermore, also LFV tree-level $Z'$ exchange is possible which could induce $\ell \to \ell_1 \ell_2 \ell_3$ processes.
The LFV couplings here arise from the mixing between the SM families and the VL family.
Although these LFV processes exist in principle,
they can easily be suppressed by certain patterns of Yukawa couplings such as $\la^{L,E}_{2} \gg \la^{L,E}_{3}, \la^{L,E}_{1}$
and gauge eigenstates which are otherwise closely aligned with the mass eigenstates.
Specific textures of the Yukawa couplings like this could be explained by flavor symmetries.

At the best fit points we find that from all possible charged LFV processes only $\br{\mu}{e\gamma}$ is close to its experimental upper bound.
Just like for $\Delta a_\mu$, the dominant contribution to $\mu \to e\gamma$ originates from the chirally enhanced $Z'$-loop with heavy leptons,
with an $\mathcal{O}\left( 10\% \right)$ cancellation arising from the $\chi$-loop contribution.

The SM boson decays may in general also be affected by mixing effects.
Models with VL leptons mixing to the SM families often affect the LFV Higgs boson decays such as
$h\to \mu\tau$, and also changes in the rates of lepton flavor conserving decays, see e.g.~\cite{Dermisek:2013gta,Altmannshofer:2016oaq,Poh:2017tfo}.
However, in the generic parameter regions of our best fit points, there is no significant contribution from the mixing to these processes.
As analytically demonstrated in Appendix~\ref{analytical_analysis} this comes about because
mixing between the SM families and the VL family are only induced by the $\U1'$ breaking scalar $\Phi$ instead of the Higgs boson.
Thus, the Higgs boson decays to SM generations are very much aligned with the SM.
The same is true for the couplings of the $Z$ boson which are very SM-like for the three SM generations.

\subsubsection{Quark Flavor Physics}
There is much literature discussing the correlation between the $\bsll$ anomalies
and $B_s$-$\ol{B}_s$ mixing since these are induced by common operators, see e.g.~\cite{DiLuzio:2017fdq}.
The recent lattice results~\cite{Aoki:2019cca,Aoki:2014nga,Gamiz:2009ku,Bazavov:2016nty}
imply that the SM contribution to $\Delta M_s$ is slightly larger than the experimental central value.
The $Z'$-boson contribution to $C^\mu_9$ also gives a constructive correction to $\Delta M_s$,
so that $\Delta M_s$ tends to deviate from the central value even more.
However, the theoretical uncertainties are large, so this is currently not the tightest
bound on the model.

We stress that unlike the case for the charged leptons, all of the SM quark families must mix in the up and/or down quark sectors
to explain the observed CKM matrix. This implies that there can be sizable NP contributions not only to $B_s$-$\ol{B}_s$ mixing
-- as commonly considered in the context of $\bsll$ anomalies --
but also to $B_d$-$\ol{B}_d$, $K$-$\ol{K}$, and $D$-$\ol{D}$ mixing.
Even if the $Z'$ contributions to the latter are much smaller at face value than those to $B_s$-$\ol{B}_s$ mixing,
the NP effects can still be significant, as also the SM contributions are further suppressed.
This is demonstrated by our two best fit points where $\Delta M_d \sim 0.6$ ps$^{-1}$ is about $1\sigma$
larger than the experimentally observed value.

In general we note that future, more precise determinations and over-constraining
of the CKM elements gives very important tests for this model, complementary to other probes.
Currently there are several tensions with current data at the $\sim1\sigma$ level, cf.\ the tables in Appendices~\ref{app:BFA} and~\ref{app:BFB}.
Very recently it has been argued that experiments may be in favor of CKM non-unitarity~\cite{Belfatto:2019swo}.
While it has to be carefully evaluated whether these hints hold up, we remark
that our model is in principle very well equipped to explain such effects.

\subsubsection{Collider Signals of Vector-Like Fermions}
The VL leptons tend to be light in order to explain $\Delta a_\mu$.
If the VL lepton is lighter than both the $Z'$ boson and the scalar $\chi$, as for example
the lightest charged VL lepton $E_1$ at point A,
it decays to a SM boson and a SM lepton,
as usually considered as a signal for VL leptons~\cite{Aad:2015dha,Dermisek:2014qca,Sirunyan:2019ofn,Bhattiprolu:2019vdu,Kumar:2015tna,Falkowski:2013jya,Ellis:2014dza}.
At point A, the lightest VL lepton $E_1$ is approximately a weak singlet
and it decays to $h\mu$, $Z\mu$, and $W\nu$ with branching fractions
of about $70\%$, $15\%$, and $15\%$, respectively.
Given the analysis of Ref.~\cite{Dermisek:2014qca}, which is based on LHC run~1 data,
the VL lepton at point A is not excluded.
The LHC Run 2 data was studied to search for a weak doublet VL lepton decaying
to a SM boson and a tau lepton in Ref.~\cite{Sirunyan:2019ofn}.
The limit from this analysis is expected to be much weaker for a weak singlet VL lepton
decaying to a muon and a SM boson.
We hope that a VL lepton of this type will be searched for by a dedicated analysis
based on LHC run~2 data.

In contrast, if the VL lepton is heavier than $\chi$ and/or $Z'$ it tends to decay to them.
For example, at point B the lightest charged VL-lepton $E_1$ predominantly decays to $\chi \mu$,
and $\chi$ subsequently decays to dimuons or di-tops, if kinematically allowed.
An expected signal in this case is
\begin{align}
 p p \to E^+_1 E^-_1 \to \mu^+ \chi (\to \mu^+\mu^-)+\mu^-  \chi (\to \mu^+\mu^-).
\end{align}
This signal is very clean with 6 muons and two pairs of dimuon resonances.
Furthermore, $(E_1, N_2)$ forms approximately a weak doublet,
such that the pair production cross section is enhanced compared to the weak singlet case.

In addition to the lightest VL leptons, also the heavier VL leptons produce distinctive signals.
These tend to decay to the lighter VL leptons, with the emission of a large number of light leptons.
For instance, at the point A, the pair produced $E_2$ gives a dramatic signal,
\begin{align}
 p p \to E_2^+ E_2^- \to Z E_1^+ + Z E_1^- \to Z \mu^+  (h/Z) + Z \mu^- (h/Z),
\end{align}
with up to 10 leptons in the final state.
These high-multiplicity lepton signals could provide a strong probe of this model.

The VL quarks are also detectable at the LHC.
Limits for VL quarks are studied in Refs.~\cite{Aaboud:2017zfn,Aaboud:2018xuw} using the LHC Run2 data,
but the decay patterns of the VL quarks in our model are much more complicated
than the ones assumed in these analyses.
Furthermore, even the lightest VL quark has a mass of $2.1(1.5)\,\mathrm{TeV}$ at the point A(B),
which is heavier than the experimental lower bound of $1.4\,\mathrm{TeV}$~\cite{Aaboud:2018xuw}.
In fact, the VL quarks are typically much heavier than both the $Z'$ or $\chi$.
They decay to a $Z'$ or $\chi$ and a SM quark with comparable branching fractions.
The signal from the pair production of the VL quarks is thus two (top) jets together with two resonance signatures.
An interesting signal arises again for the case that a boson decays to dimuons,
\begin{align}
 p p \to Q \ol{Q} \to \mathrm{jet}\ Z'(\to \mu^+\mu^-) + \mathrm{jet}\  Z' (\to \mu^+\mu^-),
\end{align}
where $Q$ is one of the VL quarks. Again, this should give very clean signals at the LHC.
Finally, note that the VL quarks can also induce missing energy signals like squarks
when both of a pair of produced VL quarks decay as $Q \to  \mathrm{jet}\ Z' (\to \nu\nu)$.

\section{Conclusion}
\label{sec-concl}

We have studied a model with a complete fourth family of vector-like fermions which are charged under a new $\U1'$ gauge symmetry.
We find parameter points at which the experimentally observed deviations in the muon anomalous magnetic moment $\Delta a_\mu$ and $\bsll$ processes
are explained without altering too much those observables that are consistent with the SM predictions.

The model can be embedded into more unified pictures, like grand unification and/or string models,
and it has a straightforward supersymmetric extension.
To this extent, it is important that all observables can be consistently explained with small $\U1'$ gauge coupling $g'$, such that
the coupling remains perturbative up to a typical GUT scale $\sim 10^{16}\,\mathrm{GeV}$.
An important consequence of demanding a resolution to $\Delta a_\mu$ is that the $\U1'$ charge assignment for the VL family
is not compatible with an $\SU5$ GUT, but instead with a Pati-Salam gauge symmetry.

In the present paper, we have displayed two good fit points
which demonstrate that this model can explain the muon anomalies
without spoiling other observables.
The explanation of the anomalies are correlated with other beyond the Standard Model predictions
for observables including lepton flavor violation, neutral meson mixing,
deviations from the SM CKM matrix and rare meson decays.
The CKM matrix in the model easily fulfills unitarity at the currently
observed level, but is in general non-unitary.
Hints for CKM non-unitarity found in a recent analysis, thus,
could easily be accommodated and would give a strong motivation to further consider this model.
Distinct signals at the LHC in $Z'\to\mu\mu$
and pair production of VL leptons and VL quarks together with
clean and distinct (resonant) multi-lepton final states
are predicted and provide important means to
test the considered parameter space.

In general, there are upper bounds on the VL fermions in order to explain the muon anomalies in this model.
It will, thus, be interesting to have a global study of how wide a parameter space is consistent with current and future experiments.
More details of our analysis and more global features of this model
will be discussed in an upcoming paper.

\section*{Acknowledgment}
The authors are grateful to R.\ Dermisek for useful discussions about this vector-like model.
The work of J.K.\ and S.R.\ is supported in part by the Department of Energy (DOE) under Award No.\ DE-SC0011726.
The work of J.K.\ is supported in part by the Grant-in-Aid for Scientific Research from the
Ministry of Education, Science, Sports and Culture (MEXT), Japan No.\ 18K13534.
The work of A.T.\ was partly supported by a postdoc fellowship of the German Academic Exchange Service~(DAAD).
A.T.\ is grateful to the Physics Department of Ohio State University and
Centro de F\'isica Te\'orica de Part\'iculas (CFTP) at Instituto Superior T\'ecnico, Lisbon
for hospitality during parts of this work.

\appendix
\section{Analytical Analysis}
\label{analytical_analysis}
We discuss the analytical expressions for the couplings in the mass basis.
We diagonalize the mass matrix in Eq.~(\ref{eq-Meg}) perturbatively
by exploiting $m_\ell  \ll \tilde{M}_\ell$,
where $m_\ell$ and $\tilde{M}_\ell$ represent the typical mass scales
of charged leptons and VL leptons, respectively.

Let us define the unitary matrices,
\begin{align}
\label{eq-U0}
 U_L^0 :=
\begin{pmatrix}
 \zv_{L_j} & \nv_{L} & \zerov \\
 0_j          & 0          & 1
\end{pmatrix},\quad
 U_R^0 :=
\begin{pmatrix}
 \zv_{E_j} & \nv_{E} & \zerov \\
 0_j          & 0          & 1
\end{pmatrix},
\end{align}
with the four-component vectors
\begin{align}
\nv_L := \frac{1}{\tilde{M}_L}
 \begin{pmatrix}
 \la^{L*}_{i} v_\Phi \\ \la_V^{L*} v_\phi
 \end{pmatrix},\quad
\nv_E := \frac{1}{\tilde{M}_E}
 \begin{pmatrix}
 \la^E_{i} v_\Phi \\ \la_V^{E} v_\phi
 \end{pmatrix},
\end{align}
and $\zv_{E_i}$, $\zv_{L_i}$, which obey the conditions
\begin{align}
\label{eq-zcond}
  \zv_{L_i}^\dagger \nv_L = \zv_{E_i}^\dagger  \nv_E = 0,\quad
 \zv_{L_i}^\dagger \zv_{L_j} = \zv_{E_i}^\dagger \zv_{E_j} = \delta_{ij}.
\end{align}
Here,
\begin{align}
\tilde{M}_L := \sqrt{ \sum_{i=1}^3 \abs{\la^L_{i}}^2 v_\Phi^2 + \abs{\la^L_V}^2 v_\phi^2 }\;,\quad
\tilde{M}_E := \sqrt{ \sum_{i=1}^3 \abs{\la^E_{i}}^2 v_\Phi^2 + \abs{\la^E_V}^2 v_\phi^2 }\;.
\end{align}
The rotated mass matrix is
\begin{align}
  \tilde{\Mcal}^e := U_R^{0\dag} \Mcal^e U_L^0  =
\begin{pmatrix}
 \tilde{y}^e_{ij} v_H    & \tilde{y}_{R_i}v_H   & 0_i \\
  \tilde{y}_{L_j}v_H           & \tilde{\la}_e v_H     & \tilde{M}_E \\
  0_j                                 & \tilde{M}_L              & \la'_e v_H
\end{pmatrix},
\end{align}
where
\begin{align}
 \tilde{y}^e_{ij} := \zv_{E_i}^\dagger \hat{y}^e  \zv_{L_j},\quad
 \tilde{y}_{R_i} :=    \zv_{E_i}^\dagger \hat{y}^e  \nv_{L}, \quad  
 \tilde{y}_{L_j} := \nv_{E}^\dagger   \hat{y}^e  \zv_{L_j},\quad
 \tilde{\la}_{e} :=     \nv_{E}^\dagger   \hat{y}^e  \nv_{L},
\end{align}
with
\begin{align}
\hat{y}^e :=
 \begin{pmatrix}
 y^e_{ij} & 0_i  \\
 0_j       & \la_e
\end{pmatrix}.
\end{align}
In this matrix,
$\tilde{y}^e_{ij}$, $\tilde{y}_{L_i}$, $\tilde{y}_{R_i}$, and $\tilde{\la}_e$ are of the order $\mathcal{O}(m_\ell/v_H)$,
while $\tilde{M}_L,\tilde{M}_E \sim\tilde{M}_\ell$. 
Here we assume $\la_e v_H \lesssim m_\mu$, in order for the muon mass to be explained without fine-tuning.
Note that $\la'_e v_H $ can be as large as the VL lepton masses if $\la'_e \sim 1$ and the VL leptons are lighter than $\sim 500\,\mathrm{GeV}$.
Hence, it cannot be treated as an expansion parameter in general.

Since the vectors $\zv'_{L_i}=[u_{L}]_{ij} \zv_{L_j}$ and $\zv'_{E_i}=[u_{E}]_{ij}\zv_{E_j}$,
for arbitrary $3\times 3$ unitary matrices $u_L$ and $u_E$, also satisfy the conditions in Eq.~\eqref{eq-zcond},
we can always find a set of vectors $\zv_{L_i}$, $\zv_{E_i}$ that diagonalize the SM Yukawa matrices,
$\tilde{y}^e_{ij} = \mathrm{diag}(y^e_1, y^e_2, y^e_3)$.
The mass matrix for the SM families then is almost diagonal
except for the mixing with the VL family induced by $\tilde{y}_L$ and $\tilde{y}_R$.
In order to explain the muon anomalies,
there should be a sizable mixing among the muon and the VL leptons,
while the mixing with the electron and tau can be suppressed in order to
avoid lepton flavor violations which are strongly constrained by experiments.
The simplest way to achieve this is by imposing the hierarchy $\la^{L,E}_2 \gg \la^{L,E}_{1,3}$.
In this case, $\tilde{y}_{L_i},  \tilde{y}_{R_i} \sim {m_\mu}/v_H$ is expected.

We can show that the unitary matrices
\begin{align}
\label{eq-U1}
U_L^1 =&\  \mathbbm{1}_{5} + \frac{1}{\tilde{M}_E}
\begin{pmatrix}
 0_{ij} & - \la'_e \tilde{y}_{L_i}^* v_H^2/\tilde{M}_L & \tilde{y}_{L_i}^* v_H \\
 \la'_e \tilde{y}_{L_j} v_H^2/\tilde{M}_L     & 0    & 0                               \\
-\tilde{y}_{L_j} v_H &0 & 0
\end{pmatrix}
+ \mathcal{O}\left(\frac{ m_\ell^2 }{\tilde{M}_\ell^2 } \right),\\
U_R^1 =&\  \mathbbm{1}_{5} + \frac{1}{\tilde{M}_L}
\begin{pmatrix}
 0_{ij} & - \la'_e \tilde{y}_{R_i} v_H^2/\tilde{M}_E  & \tilde{y}_{R_i} v_H \\
 \la'_e \tilde{y}_{R_i}^* v_H^2/\tilde{M}_E    & 0    & 0                               \\
-\tilde{y}_{R_j}^* v_H &0 & 0
\end{pmatrix}
+ \mathcal{O}\left(\frac{ m^2_{\ell}}{\tilde{M}_\ell^2 } \right),
\end{align}
block-diagonalize the mass matrix as
\begin{align}
 U_R^{1\dagger}  \tilde{\mathcal{M}}^e U_L^1  =
\mathrm{diag}
\begin{pmatrix}
\tilde{y}^e_{ij}v_H + \tilde{y}_{R_i} \tilde{y}_{L_j} \dfrac{ v_H^2}{\tilde{M}_L\tilde{M}_E}   \la'_e v_H
+ \mathcal{O}\left( \dfrac{ m^3_{\ell}}{\tilde{M}^2_\ell } \right),  &
\begin{pmatrix}
  \tilde{\la}_e v_H     & \tilde{M}_E \\
  \tilde{M}_L              & \la'_e v_H
\end{pmatrix}
\end{pmatrix}.
 \end{align}
Here, higher order corrections for the heavy states have been neglected.
The perturbative corrections to the mass matrix of the SM families are estimated as
\begin{align}
\tilde{y}_{R_i} \tilde{y}_{L_j}  \frac{v_H^2}{\tilde{M}_L\tilde{M}_E} \la'_e v_H
\sim \frac{m_\mu^2}{\tilde{M}_\ell^2}  \la'_e v_H
 =&\ 2.2 \times 10^{-5}\ \mathrm{GeV} \times \left(\frac{\la'_e v_H }{174\ \mathrm{GeV}}\right)
                                                            \left(\frac{300\ \mathrm{GeV} }{ \tilde{M}_\ell }\right)^2.
\end{align}
For typical parameters, this is much smaller than the electron mass.
Finally, we define unitary matrices $U^2_{L,R} := \mathrm{diag}\left(\mathbbm{1}_{3},  u_{L,R}\right)$
which diagonalize the mass matrix of the VL family,
\begin{align}
 u_R^\dagger
\begin{pmatrix}
  \tilde{\la}_e v_H     & \tilde{M}_E \\
  \tilde{M}_L              & \la'_e v_H
\end{pmatrix}
u_L = \mathrm{diag}\left(m_{E_1}, m_{E_2}\right).
\end{align}
Altogether, the fields in the mass basis $\hat{e}_L$, $\hat{e}_R$ can be written as
\begin{align}
\label{eq-Ue}
 e_L = U^e_L \hat{e}_L := U_L^0 U_L^1 U_L^2 \hat{e}_L,\quad
 e_R = U^e_R \hat{e}_R := U_R^0 U_R^1 U_R^2 \hat{e}_R .
\end{align}
We can now use this in order to study the 
the scalar and gauge-boson couplings in the mass basis.
Using Eqs.~(\ref{eq-U0}) and (\ref{eq-U1}), one can show that
\begin{align}
\left[ \left(U^e_R\right)^\dagger Y_e^h  U^e_L \right]_{ij} =  \tilde{y}^e_{ij} +
2\la'_e \frac{v_H^2}{\tilde{M}_L\tilde{M}_E} \tilde{y}_{R_i} \tilde{y}_{L_j}
+ \mathcal{O}\left(\frac{m_\ell^2}{\tilde{M}^2} \right),
\end{align}
and
\begin{align}
\left[ \left(U^e_L\right)^\dagger \Pfb  U^e_L     \right]_{ij} =  \delta_{ij}
   + \mathcal{O}\left(\frac{m_\ell^2}{\tilde{M}^2} \right),\quad
\left[ \left(U^e_R\right)^\dagger P_5  U^e_R    \right]_{ij} =  \mathcal{O}\left(\frac{m_\ell^2}{\tilde{M}^2} \right).
\end{align}
The sizes of the perturbative corrections are estimated as
\begin{align}
 \la'_e \frac{v_H^2}{\tilde{M}_L\tilde{M}_E} \tilde{y}_{R_i} \tilde{y}_{L_j}
\sim  \la'_e \frac{m_\mu^2}{\tilde{M}_\ell^2}
&\ =  1.2 \times 10^{-7} \times \left(\frac{\la'_e}{1.0}\right)\left( \frac{300\ \mathrm{GeV} }{ \tilde{M}_\ell}  \right)^2, \\
\frac{ m_\ell^2}{\tilde{M}_\ell^2} \lesssim \frac{m_\tau^2}{\tilde{M}_\ell^2}
    &\ = 3.5 \times 10^{-5} \times \left( \frac{300\ \mathrm{GeV} }{ \tilde{M}_\ell}  \right)^2.
\end{align}
Therefore, the Higgs boson coupling matrix is effectively diagonalized simultaneously with the mass matrix,
and the left-handed neutral current gauge interactions of the SM families in the mass basis are to a good accuracy proportional to the identity.
In principle, there are also right-handed current corrections of the $Z$/$W$ boson couplings to the SM fermions
which are tested by precision measurements of $Z$ and $W$ boson properties, see e.g.~\cite{Tanabashi:2018oca},
neutrino-nucleon scattering~\cite{Bernard:2007cf} and so on,
but their size is too small to be testable by these experiments.

The quark sector mass matrices can be diagonalized in a completely analogous fashion with the same 
conclusions. A significant difference with respect to the lepton sector can arise due to the heavy top quark mass.
The relevant expansion parameter of the perturbation then is $m_t/\tilde{M}_q$, which only
becomes $\lesssim 0.1$ if VL quark masses exceed \mbox{$\sim1.5\,\mathrm{TeV}$}.
However, in particular for the up-type quarks we may alternatively assume a hierarchy of couplings, 
$\la^Q_i v_\Phi, \la^U_i v_\Phi \ll \la^Q_V v_\phi, \la^U_V v_\phi$
which leads to the same conclusion.

\section{Details of Best Fit Point A}
\label{app:BFA}
\subsection{Input parameters}
The input parameters for the boson sector are
\begin{align}
 m_{Z'} = 494.696,\; v_\phi = 4356.63 ,\; g' = 0.266224,\; \la_\chi = 0.999975, \;  \la_\sigma = 0.997241 \;.
\end{align}
The mass matrices are
\begin{equation}
\scalebox{0.85}{$
M_e =\begin{pmatrix}
 -0.000486577& 1.51118 \cdot 10^{-6} & 1.99873 \cdot 10^{-6} & 0 & -1.14426 \cdot 10^{-6} \\
 -1.98226 \cdot 10^{-6} & -0.252886 & -0.00083412 & 0 & -179.053 \\
 -5.17686 \cdot 10^{-6} & 0.0000939331 & -1.74617 & 0 & -0.0268504 \\
 0 & 0 & 0 & -3.80325 \cdot 10^{-7} & 276.103 \\
 -0.0000358853 & 314.243 & -0.0370005 & -173.903 & -172.644 \\
\end{pmatrix}\;,
$}
\end{equation}
\begin{equation}
\scalebox{0.85}{$
M_n =
\begin{pmatrix}
 0 & 0 & 0 & -0.816618 & -4356.63 \\
 -0.0000358853 & 314.243 & -0.0370005 & -173.903 & 0.181800 \\
\end{pmatrix}\;,
$}
\end{equation}
\begin{equation}
\scalebox{0.85}{$
M_u =
 \begin{pmatrix}
 -0.00210998 & 0.118413 & 1.44892\cdot e^{-0.0260154i} & 0 & -5.43376 \\
 0.00415738 & -0.612239 & 0.701898 & 0 & -0.168839 \\
0.00168153\cdot e^{-1.90643i} & 0.132278 & -174.104 & 0 & -30.2055 \\
 0 & 0 & 0 & 0.0122855 & 4251.92 \\
 -0.0725549 & -34.1972 & -344.810 & -2091.71 & -19.5774 \\
\end{pmatrix}\;,
$}
\end{equation}
\begin{equation}
\scalebox{0.85}{$
M_d =
\begin{pmatrix}
 -0.000539915 & -0.00944158 & 0.143965\cdot e^{1.62963i} & 0 & -0.465179 \\
 0.0105179 & 0.00784947 & -1.28759 & 0 & 8.33584 \\
 0.0124618 \cdot e^{-1.98605} & 0.130429 & -2.58136 & 0 & 17.9036 \\
 0 & 0 & 0 & -0.000110584 & -2946.22 \\
 -0.0725549 & -34.1972 & -344.810 & -2091.71 & 26.9254 \\
\end{pmatrix}\;.
$}
\end{equation}
Entries involving the three families of right-handed neutrinos are omitted here
because these are suppressed by the huge Majorana mass and are irrelevant for our analysis,
see Section~\ref{sec-rslt}.

\subsection{Observables}
In Tables~\ref{tab-Al}-\ref{tab-Aq} we show results for observables at the best fit point A.
When quoted with reference, we have fitted the corresponding observable to experimental data.
Otherwise we have fitted to the tree level SM prediction which is indicated by  ``Ref.''=``SM''.
Eight observables, namely the real and imaginary parts of $C^{e,(')}_{9,10}$, are not shown
because they are at most about $10^{-10}$. 
More details on the fitting procedure will be given in our upcoming global analysis. 

For convenience we also state the extended CKM matrix at the best fit point A. It is given by 
\begin{align}
&\scalebox{0.85}{$
\hat{V}_\mathrm{CKM}= $} \\ \notag 
&\scalebox{0.85}{$
\begin{pmatrix}
 0.974475 & 0.224469 & 0.003594\cdot e^{-1.25247i}  & 0. & 0. \\
 0.224324\cdot e^{-3.14095i} & 0.973639 & 0.041311 & 0. & 0. \\
 0.008827\cdot e^{-0.385988i} & 0.040516\cdot e^{-3.12266i} & 0.999139 
& 0.001086\cdot e^{1.21990i} & 0.000008\cdot e^{3.11582i} \\
 0.000010\cdot e^{2.75752i} & 0.000044\cdot e^{0.020825i}  & 0.001084\cdot e^{-3.13970i} 
& 0.999902\cdot e^{1.22180i} & 0.013636\cdot e^{3.11681i} \\
 0.000003\cdot e^{2.77910i} & 0.000012\cdot e^{0.045064i} & 0.000287 \cdot e^{-3.11555 i} &  
0.003122\cdot e^{1.24590i} & 0.000043\cdot e^{3.14091 i} \\
 \end{pmatrix}. 
$}
\end{align}

\begin{table}[th] 
\centering 
\caption{\label{tab-Al}
Observables for charged leptons at point A. 
}
\begin{tabular}{c|ccccc}\hline 
name & value & data & Unc.& pull & Ref. \\  \hline\hline 
$m_e(m_Z)$ [GeV] $\times 10^{4}$&4.86579&4.86576&0.00049&0.068&\cite{Antusch:2013jca}  \\ 
$m_\mu(m_Z)$ [GeV]&0.102719&0.102719&0.000010&0.015&\cite{Antusch:2013jca}  \\ 
$m_\tau(m_Z)$ [GeV]&1.74617&1.74618&0.00017&0.029&\cite{Antusch:2013jca}  \\ 
\hline 
$\text{Br}\left(\mu\to e\nu\overline{\nu}\right)$&0.99997&0.99997&0.00010&0.001&SM\\ 
$\text{Br}\left(\mu^- \to e^-e^+e^-           \right)$ $\times 10^{13}$&0.000&0&7.8&0.000&\cite{Tanabashi:2018oca}  \\ 
$\text{Br}\left(\mu  \to e  \gamma \right)$ $\times 10^{13}$&3.581&0&3.3&1.093&\cite{Tanabashi:2018oca}  \\ 
\hline 
$\text{Br}\left(\tau\to e \nu\overline{\nu}\right)$&0.178510&0.178510&0.000018&0.000&SM\\ 
$\text{Br}\left(\tau\to\mu\nu\overline{\nu}\right)$&0.173612&0.173612&0.000017&0.001&SM\\ 
$\text{Br}\left(\tau^-\to e^-e^+  e^-        \right)$ $\times 10^{8}$&0.000&0&2.1&0.000&\cite{Tanabashi:2018oca}  \\ 
$\text{Br}\left(\tau^-\to e^-\mu^+e^-        \right)$ $\times 10^{8}$&0.000&0&1.2&0.000&\cite{Tanabashi:2018oca}  \\ 
$\text{Br}\left(\tau^-\to \mu^-e^+\mu^-      \right)$ $\times 10^{8}$&0.000&0&1.3&0.000&\cite{Tanabashi:2018oca}  \\ 
$\text{Br}\left(\tau^-\to \mu^-\mu^+\mu^-    \right)$ $\times 10^{8}$&3.0$\times10^{-5}$&0&1.6&0.000&\cite{Tanabashi:2018oca}  \\ 
$\text{Br}\left(\tau^-\to   e^-\mu^+\mu^-    \right)$ $\times 10^{8}$&0.000&0&2.1&0.000&\cite{Tanabashi:2018oca}  \\ 
$\text{Br}\left(\tau^-\to \mu^-  e^+  e^-    \right)$ $\times 10^{8}$&0.000&0&1.4&0.000&\cite{Tanabashi:2018oca}  \\ 
$\text{Br}\left(\tau\to e  \gamma \right)$ $\times 10^{8}$&0.000&0&2.6&0.000&\cite{Tanabashi:2018oca}  \\ 
$\text{Br}\left(\tau\to \mu\gamma \right)$ $\times 10^{8}$&2.0$\times10^{-5}$&0&3.4&0.000&\cite{Tanabashi:2018oca}  \\ 
\hline 
$\Delta a_e$ $\times 10^{13}$&-1.4$\times10^{-8}$&-8.700&3.6&2.417&\cite{Davoudiasl:2018fbb}  \\ 
$\Delta a_\mu$ $\times 10^{9}$&2.56&2.68&0.76&0.154&\cite{Tanabashi:2018oca}  \\ 
\hline\hline 
\end{tabular}  
\end{table}       
\begin{table}[th] 
\centering 
\caption{
Observables for SM bosons at point A. 
}
\begin{tabular}{c|ccccc}\hline 
name & value & data & Unc.& pull & Ref. \\  \hline\hline 
$\text{Br}\left(W^+  \to     e^+ \nu \right)$&0.10862&0.10862&0.00011&0.000&SM\\ 
$\text{Br}\left(W^+  \to   \mu^+ \nu \right)$&0.10862&0.10862&0.00011&0.000&SM\\ 
$\text{Br}\left(W^+  \to \tau^+ \nu \right)$&0.10855&0.10855&0.00011&0.000&SM\\ 
$\text{Br}\left(W   \to \text{had}   \right)$&0.652&0.666&0.025&0.550&SM\\ 
$\text{Br}\left(W^+ \to c\overline{s} \right)$&0.309&0.324&0.032&0.463&SM\\ 
\hline 
$\text{Br}\left(Z    \to    e^+    e^- \right)$ $\times 10^{2}$&3.333&3.333&0.0062&0.000&SM\\ 
$\text{Br}\left(Z    \to  \mu^+  \mu^- \right)$ $\times 10^{2}$&3.333&3.333&0.0062&0.000&SM\\ 
$\text{Br}\left(Z    \to\tau^+\tau^- \right)$ $\times 10^{2}$&3.326&3.326&0.0062&0.000&SM\\ 
$\text{Br}\left(Z   \to \text{had}   \right)$&0.676&0.677&0.025&0.014&SM\\ 
$\text{Br}\left(Z   \to u\overline{u}+c\overline{c} \right)/2$&0.1157&0.1157&0.0043&0.000&SM\\ 
$\text{Br}\left(Z  \to d\overline{d}+s\overline{s}+b\overline{b}\right)/3$&0.1483&0.1483&0.0056&0.000&SM\\ 
$\text{Br}\left(Z   \to c\overline{c} \right)$&0.1157&0.1157&0.0043&0.000&SM\\ 
$\text{Br}\left(Z   \to b\overline{b} \right)$&0.1479&0.1479&0.0056&0.000&SM\\ 
\hline 
$\text{Br}\left(Z    \to    e  \mu \right)$ $\times 10^{7}$&0.000&0&4.6&0.000&\cite{Tanabashi:2018oca}  \\ 
$\text{Br}\left(Z    \to    e \tau \right)$ $\times 10^{6}$&0.000&0&6.0&0.000&\cite{Tanabashi:2018oca}  \\ 
$\text{Br}\left(Z    \to  \mu \tau \right)$ $\times 10^{6}$&0.000&0&7.3&0.000&\cite{Tanabashi:2018oca}  \\ 
\hline 
$A_e$&0.1468&0.1468&0.0015&0.000&SM\\ 
$A_\mu$&0.147&0.147&0.015&0.000&SM\\ 
$A_\tau$&0.1468&0.1468&0.0015&0.000&SM\\ 
$A_s$&0.941&0.941&0.094&0.000&SM\\ 
$A_c$&0.6949&0.6949&0.0069&0.000&SM\\ 
$A_b$&0.9406&0.9406&0.0094&0.000&SM\\ 
\hline 
$\mu_{\mu\mu}$&0.976&0&1.3&0.751&\cite{Tanabashi:2018oca}  \\ 
$\mu_{\tau\tau}$&0.981&1.12&0.23&0.607&\cite{Tanabashi:2018oca}  \\ 
$\mu_{bb}$&0.841&0.950&0.22&0.495&\cite{Tanabashi:2018oca}  \\ 
$\mu_{\gamma\gamma}$&1.01&1.16&0.18&0.859&\cite{Tanabashi:2018oca}  \\ 
$\text{Br}\left(h    \to  e^+ e^-      \right)$ $\times 10^{3}$&4.8$\times10^{-6}$&0&1.2&0.000&\cite{Tanabashi:2018oca}  \\ 
$\text{Br}\left(h \to e    \mu \right)$ $\times 10^{4}$&0.000&0&2.1&0.000&\cite{Tanabashi:2018oca}  \\ 
$\text{Br}\left(h \to e  \tau \right)$ $\times 10^{3}$&0.000&0&4.2&0.000&\cite{Tanabashi:2018oca}  \\ 
$\text{Br}\left(h \to \mu \tau \right)$ $\times 10^{3}$&0.000&0&8.7&0.000&\cite{Tanabashi:2018oca}  \\ 
\hline\hline 
\end{tabular}  
\end{table}       
\begin{table}[th] 
\centering 
\caption{
Quark masses and CKM matrix at point A. 
}
\begin{tabular}{c|ccccc}\hline 
name & value & data & Unc.& pull & Ref. \\  \hline\hline 
$m_u(m_Z)$ [GeV] $\times 10^{3}$&1.28&1.29&0.39&0.024&\cite{Antusch:2013jca}  \\ 
$m_c(m_Z)$ [GeV]&0.623&0.627&0.019&0.186&\cite{Antusch:2013jca}  \\ 
$m_t(m_Z)$ [GeV]&171.78&171.68&1.5&0.060&\cite{Antusch:2013jca}  \\ 
$m_d(m_Z)$ [GeV] $\times 10^{3}$&2.74&2.75&0.29&0.036&\cite{Antusch:2013jca}  \\ 
$m_s(m_Z)$ [GeV] $\times 10^{3}$&53.85&54.32&2.9&0.162&\cite{Antusch:2013jca}  \\ 
$m_b(m_Z)$ [GeV]&2.85&2.85&0.026&0.036&\cite{Antusch:2013jca}  \\ 
\hline 
$\left|V_{ud}\right|$&0.97447&0.97420&0.00021&1.307&\cite{Tanabashi:2018oca}  \\ 
$\left|V_{us}\right|$&0.22447&0.22430&0.00050&0.338&\cite{Tanabashi:2018oca}  \\ 
$\left|V_{ub}\right|$ $\times 10^{3}$&3.59&3.94&0.36&0.961&\cite{Tanabashi:2018oca}  \\ 
$\left|V_{cd}\right|$&0.2243&0.2180&0.0040&1.581&\cite{Tanabashi:2018oca}  \\ 
$\left|V_{cs}\right|$&0.974&0.997&0.017&1.374&\cite{Tanabashi:2018oca}  \\ 
$\left|V_{cb}\right|$ $\times 10^{2}$&4.13&4.22&0.080&1.112&\cite{Tanabashi:2018oca}  \\ 
$\left|V_{td}\right|$ $\times 10^{3}$&8.83&8.10&0.50&1.455&\cite{Tanabashi:2018oca}  \\ 
$\left|V_{ts}\right|$ $\times 10^{2}$&4.05&3.94&0.23&0.485&\cite{Tanabashi:2018oca}  \\ 
$\left|V_{tb}\right|$&0.999&1.02&0.025&0.794&\cite{Tanabashi:2018oca}  \\ 
\hline 
$\alpha$&1.50&1.47&0.097&0.292&\cite{Tanabashi:2018oca}  \\ 
$\sin{2\beta}$&0.698&0.691&0.017&0.440&\cite{Tanabashi:2018oca}  \\ 
$\gamma$&1.25&1.28&0.081&0.382&\cite{Tanabashi:2018oca}  \\ 
\hline\hline 
\end{tabular}  
\end{table}       
\begin{table}[th] 
\centering 
\caption{\label{tab-Aq}
Observables for quarks at point A. 
}
\begin{tabular}{c|ccccc}\hline 
name & value & data & Unc.& pull & Ref. \\  \hline\hline 
$\Delta M_K$ [ps$^{-1}$] $\times 10^{3}$&4.616&5.293&2.2&0.312&\cite{Tanabashi:2018oca}  \\ 
$\epsilon_K$ $\times 10^{3}$&2.24&2.23&0.21&0.038&\cite{Tanabashi:2018oca}  \\ 
\hline 
$\Delta M_{B_d}$ [ps$^{-1}$]&0.612&0.506&0.081&1.304&\cite{Tanabashi:2018oca}  \\ 
$S_{\psi K_s}$&0.688&0.695&0.019&0.368&\cite{Amhis:2016xyh}  \\ 
\hline 
$\Delta M_{B_s}$ [ps$^{-1}$]&19.43&17.76&2.5&0.673&\cite{Tanabashi:2018oca}  \\ 
$S_{\psi \phi}$ $\times 10^{2}$&3.740&2.100&3.1&0.529&\cite{Amhis:2016xyh}  \\ 
\hline 
$\left|x^D_{12}\right|$ $\times 10^{3}$&5.3$\times10^{-5}$&0&5.0&0.000&SM\\ 
\hline 
$R_K^{\nu\nu}$&1.157&1.000&3.4&0.047&\cite{Lees:2013kla}  \\ 
$R_{K^*}^{\nu\nu}$&1.158&1.000&3.4&0.046&\cite{Lutz:2013ftz}  \\ 
$R_{B_d\to\mu\mu}$&0.860&1.509&1.4&0.457&\cite{Tanabashi:2018oca,Altmannshofer:2017wqy,Bobeth:2013uxa}  \\ 
$R_{B_s\to\mu\mu}$&0.862&0.750&0.16&0.709&\cite{Tanabashi:2018oca,Altmannshofer:2017wqy,Bobeth:2013uxa}  \\ 
$\Gamma_t$&1.49&1.41&0.17&0.485&\cite{Tanabashi:2018oca}  \\ 
$\text{Br}\left(t   \to Z q\right)$ $\times 10^{4}$&0.000&0&3.0&0.000&\cite{Tanabashi:2018oca}  \\ 
$\text{Br}\left(t   \to Z u\right)$ $\times 10^{3}$&0.000&0&1.5&0.000&\cite{Tanabashi:2018oca}  \\ 
$\text{Br}\left(t   \to Z c\right)$ $\times 10^{3}$&0.000&0&1.3&0.000&\cite{Tanabashi:2018oca}  \\ 
\hline
$\text{Re}C^\mu_9$&-0.725&-0.700&0.30&0.082&\cite{Aebischer:2019mlg}  \\ 
$\text{Im}C^\mu_9$&-7.5$\times10^{-3}$&0&0.10&0.075&\cite{Aebischer:2019mlg}  \\ 
$\text{Re}C^\mu_{10}$&0.320&0.400&0.20&0.398&\cite{Aebischer:2019mlg}  \\ 
$\text{Im}C^\mu_{10}$&3.3$\times10^{-3}$&0&0.10&0.033&\cite{Aebischer:2019mlg}  \\ 
$\text{Re}C^{'\mu}_9$&3.2$\times10^{-4}$&0&0.10&0.003&\cite{Aebischer:2019mlg}  \\ 
$\text{Im}C^{'\mu}_9$&-4.8$\times10^{-5}$&0&0.10&0.000&\cite{Aebischer:2019mlg}  \\ 
$\text{Re}C^{'\mu}_{10}$&-1.4$\times10^{-4}$&0&0.10&0.001&\cite{Aebischer:2019mlg}  \\ 
$\text{Im}C^{'\mu}_{10}$&2.1$\times10^{-5}$&0&0.10&0.000&\cite{Aebischer:2019mlg}  \\ 
$\text{Br}\left(B \to K\tau^+\tau^-\right)$ $\times 10^{3}$&1.2$\times10^{-4}$&0&1.8&0.000&\cite{TheBaBar:2016xwe}  \\ 
\hline\hline 
\end{tabular}  
\end{table}    

\clearpage
\section{Details of Best Fit Point B}
\label{app:BFB}
\subsection{Input Parameters}
The input parameters for the boson sector are
\begin{align}
 m_{Z'} = 377.090,\; v_\phi = 2915.42,\; g' = 0.305856,\; \la_\chi = 0.155278,\;\la_\sigma = 1.00000 \;.
\end{align}
The mass matrices are
\begin{equation}
\scalebox{0.85}{$
M_e =\begin{pmatrix}
 0.000486573 & 1.13824 \cdot 10^{-8} & 4.04521 \cdot 10^{-6} & 0 & -4.59216 \cdot 10^{-6} \\
 1.53811 \cdot 10^{-7} & -0.13295 & 0.0515715 & 0 & -360.594 \\
 -0.0000578552 & 0.000174906 & -1.74549 & 0 & -0.142312 \\
 0 & 0 & 0 & -0.000263151 & 1099.09 \\
 -1.26999 \cdot 10^{-6} & 337.256 & 0.0262194 & 472.383 & -174.012 \\
\end{pmatrix}\;,
$}
\end{equation}
\begin{equation}
\scalebox{0.85}{$
M_n =
\begin{pmatrix}
 0 & 0 & 0 & 2.30874 & 280.271 \\
 -1.26999 \cdot 10^{-6} & 337.256 & 0.0262194 & 472.383 & -16.3708 \\
\end{pmatrix}\;,
$}
\end{equation}
\begin{equation}
\scalebox{0.85}{$
M_u =
 \begin{pmatrix}
 0.0105038 & 0.494412 & 0.0290073\cdot e^{-0.532076i} & 0 & -0.30203 \\
 -0.00667761 & -0.384139 & -0.110669 & 0 & 7.16682 \\
 0.450446\cdot e^{-1.77965i} & 8.95794 & -171.499 & 0 & -13.1159 \\
 0 & 0 & 0 & -0.00998379 & -2915.41 \\
 -0.0540921 & -104.906 & 71.9157 & 1542.31 & -4.45885 \\
\end{pmatrix}\;,
$}
\end{equation}
\begin{equation}
\scalebox{0.85}{$
M_d =
\begin{pmatrix}
 -0.0111889 & 0.0531604 & 0.0128086\cdot e^{-0.481607i} & 0 & 0.0871297 \\
 0.00257029 & 0.00163007 & -0.0514564 & 0 & -1.52119 \\
 0.00240628\cdot e^{1.70936i} & 0.0308145 & -2.85816 & 0 & -65.9271 \\
 0 & 0 & 0 & 0.0101523 & 2901.32 \\
 -0.0540921 & -104.906 & 71.9157 & 1542.31 & 3.43706 \\
\end{pmatrix}\;.
$}
\end{equation}
\pagebreak

\subsection{Observables}
Tables \ref{tab-Bl}-\ref{tab-Bq} show the observables at the best fit point B. 
The extended CKM matrix is given by 
\begin{align}
&\scalebox{0.85}{$
\hat{V}_\mathrm{CKM}= $} \\ \notag 
&\scalebox{0.85}{$
\begin{pmatrix}
 0.974460 & 0.224533 & 0.003631\cdot e^{-1.21489i} & 0. & 0. \\
 0.224393\cdot e^{-3.14095i} & 0.973619 & 0.041392 & 0. & 0.000001\cdot e^{1.63235i} \\
 0.008717\cdot e^{-0.389908i} & 0.040627\cdot e^{-3.12277i} & 0.999136 
& 0.000623\cdot e^{-1.59708i} & 0.000022\cdot e^{1.63174i} \\
 0.000005\cdot e^{-2.00027i} & 0.000025\cdot e^{1.54988i} & 0.000622\cdot e^{-1.61053i} & 
0.999999\cdot e^{-0.066018i} & 0.000889\cdot e^{0.021202i} \\
 0.000002\cdot e^{-2.01687i} & 0.000011\cdot e^{1.52851i} & 0.000265\cdot e^{-1.63172i} & 
0.001107\cdot e^{3.05439i} & 0.000001\cdot e^{-3.14158i} \\
\end{pmatrix}. 
$}
\end{align}

\begin{table}[th] 
\centering 
\caption{\label{tab-Bl}
Observables for charged leptons at point B.}
\begin{tabular}{c|ccccc}\hline 
name & value & data & Unc.& pull & Ref. \\  \hline\hline 
$m_e(m_Z)$ [GeV] $\times 10^{4}$&4.8658&4.8658&0.00049&0.007&\cite{Antusch:2013jca}  \\ 
$m_\mu(m_Z)$ [GeV]&0.102719&0.102719&0.000010&0.045&\cite{Antusch:2013jca}  \\ 
$m_\tau(m_Z)$ [GeV]&1.7462&1.7462&0.00017&0.039&\cite{Antusch:2013jca}  \\ 
\hline 
$\text{Br}\left(\mu\to e\nu\overline{\nu}\right)$&0.99995&0.99997&0.00010&0.229&SM\\ 
$\text{Br}\left(\mu^- \to e^-e^+e^-           \right)$ $\times 10^{13}$&0.000&0&7.8&0.000&\cite{Tanabashi:2018oca}  \\ 
$\text{Br}\left(\mu  \to e  \gamma \right)$ $\times 10^{13}$&2.100&0&3.3&0.641&\cite{Tanabashi:2018oca}  \\ 
\hline 
$\text{Br}\left(\tau\to e \nu\overline{\nu}\right)$&0.178510&0.178510&0.000018&0.000&SM\\ 
$\text{Br}\left(\tau\to\mu\nu\overline{\nu}\right)$&0.173608&0.173612&0.000017&0.229&SM\\ 
$\text{Br}\left(\tau^-\to e^-e^+  e^-        \right)$ $\times 10^{8}$&0.000&0&2.1&0.000&\cite{Tanabashi:2018oca}  \\ 
$\text{Br}\left(\tau^-\to e^-\mu^+e^-        \right)$ $\times 10^{8}$&0.000&0&1.2&0.000&\cite{Tanabashi:2018oca}  \\ 
$\text{Br}\left(\tau^-\to \mu^-e^+\mu^-      \right)$ $\times 10^{8}$&0.000&0&1.3&0.000&\cite{Tanabashi:2018oca}  \\ 
$\text{Br}\left(\tau^-\to \mu^-\mu^+\mu^-    \right)$ $\times 10^{8}$&7.4$\times10^{-3}$&0&1.6&0.005&\cite{Tanabashi:2018oca}  \\ 
$\text{Br}\left(\tau^-\to   e^-\mu^+\mu^-    \right)$ $\times 10^{8}$&0.000&0&2.1&0.000&\cite{Tanabashi:2018oca}  \\ 
$\text{Br}\left(\tau^-\to \mu^-  e^+  e^-    \right)$ $\times 10^{8}$&0.000&0&1.4&0.000&\cite{Tanabashi:2018oca}  \\ 
$\text{Br}\left(\tau\to e  \gamma \right)$ $\times 10^{8}$&0.000&0&2.6&0.000&\cite{Tanabashi:2018oca}  \\ 
$\text{Br}\left(\tau\to \mu\gamma \right)$ $\times 10^{8}$&4.9$\times10^{-2}$&0&3.4&0.014&\cite{Tanabashi:2018oca}  \\ 
\hline 
$\Delta a_e$ $\times 10^{13}$&0.000&-8.700&3.6&2.417&\cite{Davoudiasl:2018fbb}  \\ 
$\Delta a_\mu$ $\times 10^{9}$&2.43&2.68&0.76&0.332&\cite{Tanabashi:2018oca}  \\ 
\hline\hline 
\end{tabular}  
\end{table}      
\begin{table}[th] 
\centering 
\caption{Observables for SM bosons at point B.}
\begin{tabular}{c|ccccc}\hline 
name & value & data & Unc.& pull & Ref. \\  \hline\hline 
$\text{Br}\left(W^+  \to     e^+ \nu \right)$&0.10862&0.10862&0.00011&0.000&SM\\ 
$\text{Br}\left(W^+  \to   \mu^+ \nu \right)$&0.10862&0.10862&0.00011&0.023&SM\\ 
$\text{Br}\left(W^+  \to \tau^+ \nu \right)$&0.10855&0.10855&0.00011&0.000&SM\\ 
$\text{Br}\left(W   \to \text{had}   \right)$&0.652&0.666&0.025&0.550&SM\\ 
$\text{Br}\left(W^+ \to c\overline{s} \right)$&0.309&0.324&0.032&0.464&SM\\ 
\hline 
$\text{Br}\left(Z    \to    e^+    e^- \right)$ $\times 10^{2}$&3.333&3.333&0.0062&0.000&SM\\ 
$\text{Br}\left(Z    \to  \mu^+  \mu^- \right)$ $\times 10^{2}$&3.333&3.333&0.0062&0.000&SM\\ 
$\text{Br}\left(Z    \to\tau^+\tau^- \right)$ $\times 10^{2}$&3.326&3.326&0.0062&0.000&SM\\ 
$\text{Br}\left(Z   \to \text{had}   \right)$&0.676&0.677&0.025&0.014&SM\\ 
$\text{Br}\left(Z   \to u\overline{u}+c\overline{c} \right)/2$&0.1157&0.1157&0.0043&0.000&SM\\ 
$\text{Br}\left(Z  \to d\overline{d}+s\overline{s}+b\overline{b}\right)/3$&0.1483&0.1483&0.0056&0.000&SM\\ 
$\text{Br}\left(Z   \to c\overline{c} \right)$&0.1157&0.1157&0.0043&0.000&SM\\ 
$\text{Br}\left(Z   \to b\overline{b} \right)$&0.1479&0.1479&0.0056&0.000&SM\\ 
\hline 
$\text{Br}\left(Z    \to    e  \mu \right)$ $\times 10^{7}$&0.000&0&4.6&0.000&\cite{Tanabashi:2018oca}  \\ 
$\text{Br}\left(Z    \to    e \tau \right)$ $\times 10^{6}$&0.000&0&6.0&0.000&\cite{Tanabashi:2018oca}  \\ 
$\text{Br}\left(Z    \to  \mu \tau \right)$ $\times 10^{6}$&0.000&0&7.3&0.000&\cite{Tanabashi:2018oca}  \\ 
\hline 
$A_e$&0.1468&0.1468&0.0015&0.000&SM\\ 
$A_\mu$&0.147&0.147&0.015&0.000&SM\\ 
$A_\tau$&0.1468&0.1468&0.0015&0.000&SM\\ 
$A_s$&0.941&0.941&0.094&0.000&SM\\ 
$A_c$&0.6949&0.6949&0.0069&0.000&SM\\ 
$A_b$&0.9406&0.9406&0.0094&0.000&SM\\ 
\hline 
$\mu_{\mu\mu}$&0.977&0&1.3&0.752&\cite{Tanabashi:2018oca}  \\ 
$\mu_{\tau\tau}$&0.981&1.12&0.23&0.606&\cite{Tanabashi:2018oca}  \\ 
$\mu_{bb}$&0.843&0.950&0.22&0.488&\cite{Tanabashi:2018oca}  \\ 
$\mu_{\gamma\gamma}$&1.00&1.16&0.18&0.863&\cite{Tanabashi:2018oca}  \\ 
$\text{Br}\left(h    \to  e^+ e^-      \right)$ $\times 10^{3}$&4.8$\times10^{-6}$&0&1.2&0.000&\cite{Tanabashi:2018oca}  \\ 
$\text{Br}\left(h \to e    \mu \right)$ $\times 10^{4}$&0.000&0&2.1&0.000&\cite{Tanabashi:2018oca}  \\ 
$\text{Br}\left(h \to e  \tau \right)$ $\times 10^{3}$&0.000&0&4.2&0.000&\cite{Tanabashi:2018oca}  \\ 
$\text{Br}\left(h \to \mu \tau \right)$ $\times 10^{3}$&0.000&0&8.7&0.000&\cite{Tanabashi:2018oca}  \\ 
\hline\hline 
\end{tabular}  
\end{table}    
\begin{table}[th] 
\centering 
\caption{Quark masses and CKM matrix  at the point B.}
\begin{tabular}{c|ccccc}\hline 
name & value & data & Unc.& pull & Ref. \\  \hline\hline 
$m_u(m_Z)$ [GeV] $\times 10^{3}$&1.28&1.29&0.39&0.031&\cite{Antusch:2013jca}  \\ 
$m_c(m_Z)$ [GeV]&0.629&0.627&0.019&0.096&\cite{Antusch:2013jca}  \\ 
$m_t(m_Z)$ [GeV]&171.52&171.68&1.5&0.112&\cite{Antusch:2013jca}  \\ 
$m_d(m_Z)$ [GeV] $\times 10^{3}$&2.74&2.75&0.29&0.032&\cite{Antusch:2013jca}  \\ 
$m_s(m_Z)$ [GeV] $\times 10^{3}$&54.33&54.32&2.9&0.002&\cite{Antusch:2013jca}  \\ 
$m_b(m_Z)$ [GeV]&2.85&2.85&0.026&0.052&\cite{Antusch:2013jca}  \\ 
\hline 
$\left|V_{ud}\right|$&0.97446&0.97420&0.00021&1.236&\cite{Tanabashi:2018oca}  \\ 
$\left|V_{us}\right|$&0.22453&0.22430&0.00050&0.467&\cite{Tanabashi:2018oca}  \\ 
$\left|V_{ub}\right|$ $\times 10^{3}$&3.63&3.94&0.36&0.859&\cite{Tanabashi:2018oca}  \\ 
$\left|V_{cd}\right|$&0.2244&0.2180&0.0040&1.598&\cite{Tanabashi:2018oca}  \\ 
$\left|V_{cs}\right|$&0.974&0.997&0.017&1.375&\cite{Tanabashi:2018oca}  \\ 
$\left|V_{cb}\right|$ $\times 10^{2}$&4.14&4.22&0.080&1.010&\cite{Tanabashi:2018oca}  \\ 
$\left|V_{td}\right|$ $\times 10^{3}$&8.72&8.10&0.50&1.233&\cite{Tanabashi:2018oca}  \\ 
$\left|V_{ts}\right|$ $\times 10^{2}$&4.06&3.94&0.23&0.533&\cite{Tanabashi:2018oca}  \\ 
$\left|V_{tb}\right|$&0.999&1.02&0.025&0.795&\cite{Tanabashi:2018oca}  \\ 
\hline 
$\alpha$&1.54&1.47&0.097&0.640&\cite{Tanabashi:2018oca}  \\ 
$\sin{2\beta}$&0.704&0.691&0.017&0.768&\cite{Tanabashi:2018oca}  \\ 
$\gamma$&1.21&1.28&0.081&0.845&\cite{Tanabashi:2018oca}  \\ 
\hline\hline 
\end{tabular}  
\end{table}        
\begin{table}[th] 
\centering 
\caption{\label{tab-Bq} 
Observables for quarks at point B.}
\begin{tabular}{c|ccccc}\hline 
name & value & data & Unc.& pull & Ref. \\  \hline\hline 
$\Delta M_K$ [ps$^{-1}$] $\times 10^{3}$&6.015&5.293&2.2&0.333&\cite{Tanabashi:2018oca}  \\ 
$\epsilon_K$ $\times 10^{3}$&2.21&2.23&0.21&0.081&\cite{Tanabashi:2018oca}  \\ 
\hline 
$\Delta M_{B_d}$ [ps$^{-1}$]&0.599&0.506&0.081&1.139&\cite{Tanabashi:2018oca}  \\ 
$S_{\psi K_s}$&0.686&0.695&0.019&0.475&\cite{Amhis:2016xyh}  \\ 
\hline 
$\Delta M_{B_s}$ [ps$^{-1}$]&19.81&17.76&2.5&0.826&\cite{Tanabashi:2018oca}  \\ 
$S_{\psi \phi}$ $\times 10^{2}$&3.627&2.100&3.1&0.493&\cite{Amhis:2016xyh}  \\ 
\hline 
$\left|x^D_{12}\right|$ $\times 10^{3}$&2.0$\times10^{-2}$&0&5.0&0.004&SM\\ 
\hline 
$R_K^{\nu\nu}$&1.133&1.000&3.4&0.040&\cite{Lees:2013kla}  \\ 
$R_{K^*}^{\nu\nu}$&1.133&1.000&3.4&0.039&\cite{Lutz:2013ftz}  \\ 
$R_{B_d\to\mu\mu}$&0.877&1.509&1.4&0.445&\cite{Tanabashi:2018oca,Altmannshofer:2017wqy,Bobeth:2013uxa}  \\ 
$R_{B_s\to\mu\mu}$&0.864&0.750&0.16&0.722&\cite{Tanabashi:2018oca,Altmannshofer:2017wqy,Bobeth:2013uxa}  \\ 
$\Gamma_t$&1.49&1.41&0.17&0.485&\cite{Tanabashi:2018oca}  \\ 
$\text{Br}\left(t   \to Z q\right)$ $\times 10^{4}$&0.000&0&3.0&0.000&\cite{Tanabashi:2018oca}  \\ 
$\text{Br}\left(t   \to Z u\right)$ $\times 10^{3}$&0.000&0&1.5&0.000&\cite{Tanabashi:2018oca}  \\ 
$\text{Br}\left(t   \to Z c\right)$ $\times 10^{3}$&0.000&0&1.3&0.000&\cite{Tanabashi:2018oca}  \\ 
\hline
$\text{Re}C^\mu_9$&-0.571&-0.700&0.30&0.431&\cite{Aebischer:2019mlg}  \\ 
$\text{Im}C^\mu_9$&-1.0$\times10^{-2}$&0&0.10&0.103&\cite{Aebischer:2019mlg}  \\ 
$\text{Re}C^\mu_{10}$&0.316&0.400&0.20&0.421&\cite{Aebischer:2019mlg}  \\ 
$\text{Im}C^\mu_{10}$&5.7$\times10^{-3}$&0&0.10&0.057&\cite{Aebischer:2019mlg}  \\ 
$\text{Re}C^{'\mu}_9$&-2.5$\times10^{-4}$&0&0.10&0.002&\cite{Aebischer:2019mlg}  \\ 
$\text{Im}C^{'\mu}_9$&1.9$\times10^{-4}$&0&0.10&0.002&\cite{Aebischer:2019mlg}  \\ 
$\text{Re}C^{'\mu}_{10}$&1.4$\times10^{-4}$&0&0.10&0.001&\cite{Aebischer:2019mlg}  \\ 
$\text{Im}C^{'\mu}_{10}$&-1.1$\times10^{-4}$&0&0.10&0.001&\cite{Aebischer:2019mlg}  \\ 
$\text{Br}\left(B \to K\tau^+\tau^-\right)$ $\times 10^{3}$&1.2$\times10^{-4}$&0&1.8&0.000&\cite{TheBaBar:2016xwe}  \\ 
\hline\hline 
\end{tabular}  
\end{table}    
\clearpage

{\small
\bibliographystyle{JHEP}
\bibliography{reference_vectorlike}
}

\end{document}